\documentclass[
reprint,
superscriptaddress,
amsmath,amssymb,
aps,
prb,
twocolumn,
floatfix
]{revtex4-2}

\usepackage{bm}
\usepackage{physics}
\usepackage{dsfont}

\usepackage{graphicx}
\graphicspath{{./}}
\usepackage{subfigure}
\usepackage{epsfig}
\usepackage{placeins}

\usepackage{dcolumn}
\usepackage{enumitem}

\usepackage{tikz}
\usepackage{xcolor}
\usepackage{lipsum}
\usepackage[normalem]{ulem}
\usepackage[breaklinks]{hyperref}
\hypersetup{colorlinks=true, linkcolor=blue, citecolor=blue, filecolor=blue, urlcolor=blue}
\usepackage{cleveref}

\makeatletter
\newcommand*{\rom}[1]{\expandafter\@slowromancap\romannumeral #1@}
\makeatother

\definecolor{xqred}{RGB}{210, 70, 90}

\begin{document}
\title{
Restoring quantum neural network trainability through reusable calibration
}

\author{Ruizhe Shen}
\email{e0554228@u.nus.edu}
\altaffiliation{These authors contributed equally and are listed alphabetically.}
\affiliation{Department of Physics, National University of Singapore, Singapore 117551}
\author{Xianquan Yan}
\email{yanx@u.nus.edu}
\altaffiliation{These authors contributed equally and are listed alphabetically.}
\affiliation{Department of Physics, National University of Singapore, Singapore 117551}
\affiliation{Department of Computer Science, National University of Singapore, Singapore 117417}

\author{Kenji Kawaguchi}
\affiliation{Department of Computer Science, National University of Singapore, Singapore 117417}

\author{Ching Hua Lee}
\email{phylch@nus.edu.sg}
\affiliation{Department of Physics, National University of Singapore, Singapore 117551}

\date{\today}

\begin{abstract}
{Quantum neural networks (QNNs) rely on measured quantum observables to guide optimization, but hardware noise can suppress the parameter-dependent signals essential for learning. Here, we develop a structure-preserving error-mitigation framework that improves QNN trainability through reusable calibration of the quantum circuit's effective noise response. By calibrating a reference circuit with the same gate architecture as the trainable circuit, we construct regularized estimators of local quantum features that remain fixed throughout training. We quantify how calibration errors and response mismatches affect the recovery of both observables and their derivatives. Numerical simulations demonstrate enhanced loss-landscape contrast, faster convergence, and improved classification performance, and an IBM Quantum experiment confirms improved closed-loop variational optimization under hardware noise. We further demonstrate calibration reuse in a modular QNN comprising 25 independent four-qubit channels, corresponding to 100 qubits in aggregate. Our results establish reusable feature calibration as a practical approach to improving noise-limited quantum machine learning without repeated calibration at every training step.}
\end{abstract}

\maketitle
%\onecolumngrid

\section{Introduction}\label{sec0}

Quantum neural networks (QNNs) rely on measurements of parameterized quantum circuits to extract features that define the learning objective and guide classical optimization \cite{mitarai2018quantum,farhi2018classification,benedetti2019parameterized,beer2020training,biamonte2017quantum,cerezo2022challenges}. However, hardware noise can distort both the measured features and their dependence on circuit parameters, reducing the ability to distinguish different parameter configurations and weakening the gradients that drive learning \cite{wang2021noise,arrasmith2021effect,larocca2024review}. Because QNN training continuously updates the variational parameters while processing different input data, effective error mitigation must remain reliable across a family of evolving quantum circuits, rather than correcting the output of a single fixed circuit.

Quantum error mitigation offers a means of recovering observables from noisy quantum processors without requiring full quantum error correction \cite{temme2017error,endo2018practical,cai2023quantum,endo2021hybrid,song2019mitigation,kim2023utility,li2026practical}. Existing approaches address different sources of error. Readout calibration corrects measurement-assignment errors \cite{bravyi2021mitigating,nation2021scalable,maciejewski2020detector,nachman2020unfolding,chen2023aklt,shen2025skin}, while zero-noise extrapolation and probabilistic error cancellation mitigate errors accumulated during circuit execution at the cost of additional quantum sampling \cite{kandala2019extending,vandenberg2023probabilistic,sun2021realistic,kim2023scalable,ferracin2024efficient,shen2025robust}. Clifford data regression instead constructs corrections using classically tractable reference circuits \cite{czarnik2021error,lowe2021unified,bultrini2023unifying}. 
{Although several error-mitigation methods allow corrections to be reused across circuit evaluations \cite{strikis2021learning,korolev2026machine,robbiati2026realtime}, their application to QNN training presents additional challenges. In particular, readout mitigation does not correct errors accumulated during circuit execution, while methods requiring auxiliary or noise-scaled circuit evaluations can introduce substantial overhead when applied repeatedly throughout optimization. For QNNs, the central challenge is therefore to mitigate circuit-induced distortions of the measured quantum features while maintaining an efficient and consistent correction throughout training. This motivates the development of a structure-preserving calibration strategy that exploits the repeated circuit architecture to recover the feature information needed for learning, without requiring additional mitigation circuits or recalibration at every parameter update.}

{To address this challenge, we develop a structure-preserving error-mitigation framework that directly corrects the local Pauli-$Z$ observables used as quantum features in QNN training. The method constructs a reference circuit that approximates the identity while preserving the gate structure, depth, and connectivity of the trainable quantum circuit \cite{shen2025structure}. Measurements of computational-basis inputs are then used to characterize its effective noise response and construct a regularized linear estimator, which is applied to the measured features throughout training without repeated calibration at each parameter update.}

{An important advantage of this approach is that the calibration complexity is determined by the size of an individual quantum channel rather than the total network width. For a channel containing $N$ qubits, calibration requires $2^N$ computational-basis input settings to construct a $2^N\times 2^N$ response matrix. When different channels exhibit compatible effective noise responses, the same calibration can be reused across the network. We illustrate this capability using a modular QNN comprising 25 independent four-qubit channels, corresponding to 100 qubits in aggregate, while retaining a four-qubit calibration block. We further investigate the impact of reusable feature correction on QNN training through numerical classification benchmarks and demonstrate its effectiveness in closed-loop variational optimization on IBM Quantum hardware. Together, these results connect the theoretical conditions for calibration transfer with practical improvements in noise-limited variational learning.}

\section{Results}

\begin{figure*}[!t]
    \centering
    \includegraphics[width=\linewidth]{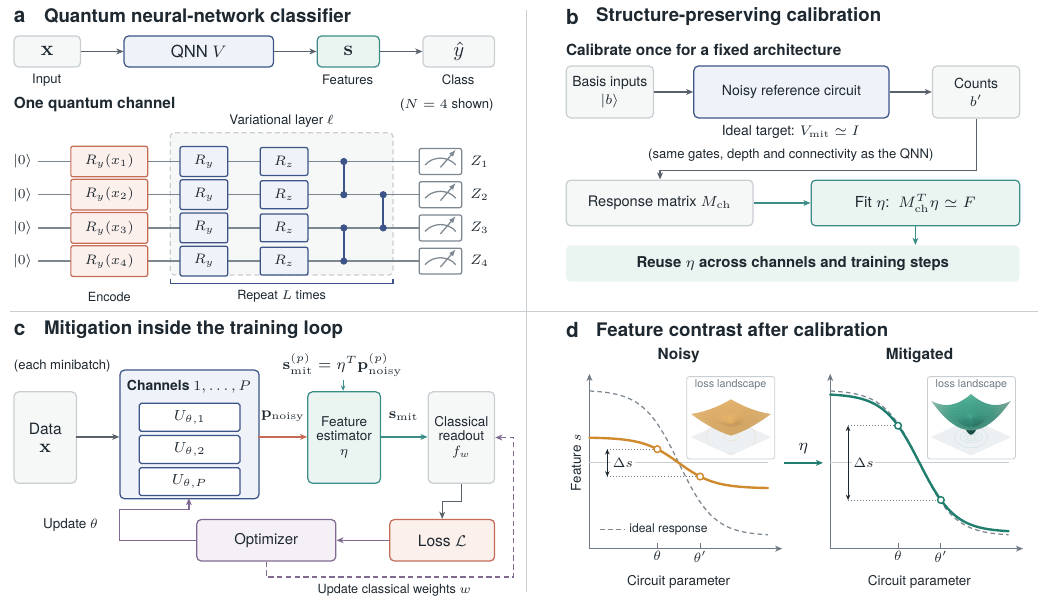}
    \caption{{Reusable calibration corrects the quantum features entering QNN training.
(a) A classical input $\mathbf{x}$ is encoded by $R_y(x_q)$ rotations. The illustrative four-qubit channel $V$ contains $L$ layers of trainable $R_y$ and $R_z$ rotations followed by even- and odd-bond controlled-$Z$ (CZ) gates; connected filled dots denote CZ gates. Local Pauli-$Z$ measurements provide the feature vector $\mathbf{s}$ used to predict the class $\hat y$.
(b) Computational-basis inputs $\ket b$ and counts of output bitstrings $b'$ characterize an identity-target reference, $V_{\mathrm{mit}}\simeq I$, with the same gate structure, depth and connectivity as the QNN channel. Its response matrix $M_{\mathrm{ch}}\equiv M_{\mathrm{mit}}$ determines a regularized estimator $\eta$ satisfying $M_{\mathrm{ch}}^T\eta\simeq F$, where $F_{bq}=\langle b|Z_q|b\rangle$. The estimator can be reused across parameter updates and channels with compatible effective noise responses.
(c) Parallel channels $U_{\theta,1},\ldots,U_{\theta,P}$ produce noisy bitstring probabilities $\mathbf{p}_{\mathrm{noisy}}$. Applying the fixed estimator gives $\mathbf{s}_{\mathrm{mit}}=\eta^T\mathbf{p}_{\mathrm{noisy}}$ for each channel, and the combined features enter the classical readout $f_w$ and loss $\mathcal{L}$. Solid and dashed purple feedback paths indicate updates of quantum parameters $\theta$ and classical parameters $w$, respectively; $w$ corresponds to $\mathcal{W}$ in the text.
(d) Schematic correction of a measured feature during optimization. Gray dashed curves show the ideal response to a circuit parameter; orange and teal curves show attenuation by noise and correction with a fixed estimator, respectively. Vertical arrows indicate the feature difference $\Delta s=s(\theta)-s(\theta')$ between two values $\theta$ and $\theta'$ of one circuit parameter; these parameter values are the same in both plots. Under a compatible effective response, calibration recovers the feature changes that contribute to the loss and its parameter derivatives. Insets show the corresponding schematic loss landscapes on a common loss scale as conceptual illustrations.}}
    \label{fig:fig1}
\end{figure*}

\subsection{Error-mitigated quantum neural-network architecture}
{We consider a modular QNN classifier whose quantum layer consists of $P$ parallel channels of $N$ qubits each [Fig.~\ref{fig:fig1}(c)]. Panel (a) shows the circuit of one illustrative four-qubit channel and the flow from its input to measured features and a class prediction.} The notation is summarized in Appendix~\ref{sec:glossary}. A trainable linear layer maps an input $\mathbf{x}\in\mathbb{R}^{K}$ to an $NP$-dimensional latent vector, which is divided into $P$ channel inputs,
\begin{equation}
\mathbf{x}_{\mathrm{latent}}
    ={\begin{bmatrix}\mathbf{x}^{(1)}\\ \vdots\\ \mathbf{x}^{(P)}\end{bmatrix}},
    \qquad \mathbf{x}^{(\alpha)}\in\mathbb{R}^{N}.
    \label{eq:latent_channels}
\end{equation}

Each subvector is processed by one $N$-qubit quantum channel. Starting from $\ket{0}^{\otimes N}$, the $\alpha$-th channel encodes the input through single-qubit rotations, a standard form of angle encoding used in variational quantum learning models \cite{mitarai2018quantum,schuld2019quantum,schuld2020circuit,schuld2021encoding,perezsalinas2020reupload},
\begin{equation}
V_{\mathrm{enc}}\left(\mathbf{x}^{(\alpha)}\right)
=
\bigotimes_{q=1}^{N}
R_y\left(x_q^{(\alpha)}\right),
\label{eq:qnn_encoding}
\end{equation}
followed by a trainable circuit,
\begin{equation}
V\left(\mathbf{x}^{(\alpha)},\boldsymbol{\theta}^{(\alpha)}\right)
=
V_{\mathrm{train}}\left(\boldsymbol{\theta}^{(\alpha)}\right)
V_{\mathrm{enc}}\left(\mathbf{x}^{(\alpha)}\right),
\label{eq:vqnn}
\end{equation}
where $V_{\mathrm{train}}$ contains parameterized single-qubit rotations and a fixed entangling pattern. {Figure~\ref{fig:fig1} illustrates a CZ-chain architecture. The retained classification runs in Figs.~\ref{fig:loss} and~\ref{fig:data2} use the full-CX implementation specified in Appendix~\ref{app:qnn_circuit}, within the hardware-efficient ansatz paradigm} \cite{kandala2017hardware,sim2019expressibility,cerezo2021variational,holmes2021connecting,peruzzo2014variational,mcclean2016theory,shen2026crossovers,chen2026largeperiod}. {The channel circuit is labeled $V$ in Fig.~\ref{fig:fig1}(a) and $U_{\theta,\alpha}$ for channel $\alpha$ in panel (c).} Further circuit details are given in Appendix~\ref{app:qnn_circuit}.

The channel output is the local Pauli-$Z$ feature vector,
\begin{equation}
\mathbf{s}^{(\alpha)}
    =\big[\langle Z^{(\alpha)}_1\rangle,\ldots,\langle Z^{(\alpha)}_N\rangle\big]^T,
    \label{eq:channel_feature_vector}
\end{equation}
which is estimated from the bitstring probabilities measured at the output of channel $\alpha$~\cite{chen2024direct,shen2026landau,ng2026knotted}. The features from all channels are concatenated and passed to a classical postprocessing layer, which produces the predicted class probabilities $\mathbf{f}=[f_1,\ldots,f_C]^T$~\cite{havlicek2019supervised,abbas2021power,cong2019quantum}. The quantum and classical parameters are trained by minimizing the standard cross-entropy loss,
\begin{equation}
\mathcal{L}(\boldsymbol{\theta},\mathcal{W})
=
-
\sum_{k=1}^{C}
y_k
\ln
f_k(\mathbf{x},\boldsymbol{\theta},\mathcal{W}),
\label{eq:loss}
\end{equation}
where $y_k$ is the one-hot target label and $\mathcal{W}$ collects the trainable classical weights and biases. Appendix~\ref{Method} gives the complete model structure. Hardware noise directly distorts this learning problem because the measured features entering the loss can be contracted by decoherence, gate errors, and readout errors \cite{mcclean2018barren,wang2021noise,cerezo2021cost,larocca2024review}. {A simple illustration is a parameter-independent contraction of the measured features,}
\begin{equation}
\mathbf{s}^{\rm noisy}
\simeq
\lambda\mathbf{s}^{\rm ideal},
\qquad
0<\lambda<1,
\label{eq:noisy_feature_simple}
\end{equation}
{Note that this is a restricted response model for illustration, not an assumption of the implemented estimator: a general channel can attenuate different observables differently~\cite{shen2026averagedspt} and can produce parameter-dependent mixing. Moreover, the full loss gradient also depends on the classical readout evaluated at the changed features.}

To address this problem, we introduce a structure-preserving mitigation protocol that calibrates the effective noise response of a fixed QNN architecture. Rather than characterizing each microscopic noise source or reconstructing a detailed device-level noise model \cite{endo2018practical,cai2023quantum,takagi2022fundamental}, we keep the gate set, connectivity, and layer structure fixed during both calibration and training. {Reference circuits with known ideal outputs provide calibration labels without reconstructing a microscopic noise model \cite{shen2025structure,czarnik2021error,strikis2021learning,urbanek2021mitigating}. Here, the reference is used to determine fixed estimators of the classifier features, whose transfer through the training family is analyzed below.}

{Figure~\ref{fig:fig1}(b) separates calibration from the subsequent training loop. We construct a reference circuit $V_{\mathrm{mit}}$ with the same gate structure, depth and connectivity as one $N$-qubit QNN channel and fit its parameters classically so that $V_{\mathrm{mit}}\approx I$.} Because the ideal reference leaves every input unchanged, any deviation between its input and output bitstrings probes the effective noise of the shared circuit structure. We verify each fit through the Frobenius-norm distance between the fitted unitary and the identity [Appendix~\ref{sec:mitigation}]. We obtain the raw reference response matrix by preparing each computational-basis state $\{\ket{b}\}_{b=0}^{2^N-1}$ and measuring the output bitstring $b'$
\begin{equation}
    (M_{\mathrm{ref}})_{b',b}
    =
    \Pr(b'\mid b).
    \label{eq:mch}
\end{equation}
For a verified identity reference, the calibration response is simply $M_{\mathrm{mit}}=M_{\mathrm{ref}}$. More generally, if the ideal reference maps basis states to known basis states, define its permutation matrix $\Pi_{b'b}=|\langle b'|V_{\mathrm{mit}}|b\rangle|^2$ and relabel the response columns:
\begin{equation}
M_{\mathrm{mit}}=M_{\mathrm{ref}}\Pi^T.
\label{eq:reference_relabel}
\end{equation}
This removes the known ideal reference action from the calibration labels, so that $M_{\mathrm{mit}}$ maps ideal output distributions to noisy ones. A reference that coherently mixes basis states requires a different calibration design.

The classifier does not use the full bitstring distribution of a channel, but only its local Pauli-$Z$ features. These features are linear in the output probabilities. With $F\in\{-1,+1\}^{2^N\times N}$ collecting the eigenvalues $F_{bq}=\langle b|Z_q|b\rangle$, one channel gives $\mathbf{s}^{\mathrm{ideal}}=F^T\mathbf{p}^{\mathrm{ideal}}$. We therefore correct the features directly with a $2^N\times N$ feature estimator $\eta$, which maps the noisy probabilities of a channel to $\mathbf{s}^{\mathrm{mit}}=\eta^T\mathbf{p}^{\mathrm{noisy}}$. Under $\mathbf{p}^{\mathrm{noisy}}\simeq M_{\mathrm{mit}}\mathbf{p}^{\mathrm{ideal}}$, this estimate reproduces the ideal features provided that
\begin{equation}
    M_{\mathrm{mit}}^{T}\eta
    \simeq
    F .
    \label{eq:constraint_eta}
\end{equation}
We solve Eq.~\eqref{eq:constraint_eta} by Tikhonov-regularized least squares, which limits the amplification of finite-shot calibration noise [Appendix~\ref{sec:mitigation}]. The mitigation is thus a regularized feature estimator rather than an inverse of the full noise process, and it requires neither reconstruction of the output quantum state nor full error correction. For a QNN with $P$ parallel channels, the same estimator is applied independently to each channel,
\begin{equation}
\mathbf{s}^{\mathrm{mit}}
    ={\begin{bmatrix}[\mathbf{s}^{(1)}]^{\mathrm{mit}}\\ \vdots\\ [\mathbf{s}^{(P)}]^{\mathrm{mit}}\end{bmatrix}}.
    \label{eq:mmit_product}
\end{equation}
When the channels share a compatible effective response, calibration is performed once for a single $N$-qubit circuit block and reused across all $P$ channels, keeping the response dimension at $2^N\times2^N$. Shared topology defines the calibration family but does not by itself guarantee this response condition, and channels with different responses require separate small-block calibrations. {Within the training loop [Fig.~\ref{fig:fig1}(c)], each channel supplies noisy bitstring probabilities to the fixed estimator. The corrected feature vectors are concatenated for the classical readout, while the optimizer updates the quantum and classical parameters and leaves $\eta$ unchanged. Calibration reuse therefore removes the need to refit this estimator at each parameter update, provided the effective response remains compatible.}

{Figure~\ref{fig:fig1}(d) summarizes how feature correction can affect optimization. Noise can suppress the parameter dependence of the measured features and hence reduce the loss variation available to the optimizer. For fixed classical weights, recovering the feature vectors at two parameter points recovers their ideal loss difference, because the same classical readout acts on those features. Recovering their parameter derivatives also restores the corresponding loss gradient. In the main plots, orange and teal curves show one measured feature before and after correction, and $\Delta s$ marks its change between two parameter values $\theta$ and $\theta'$. The insets show the corresponding loss landscapes, where the recovered feature contrast appears as a deeper loss basin. Both the curves and the landscapes are schematic. The recovery conditions are derived in Appendix~\ref{supp:observable_recovery}.}

\begin{figure*}[!t]
	\centering
	\includegraphics[width=0.99\linewidth]{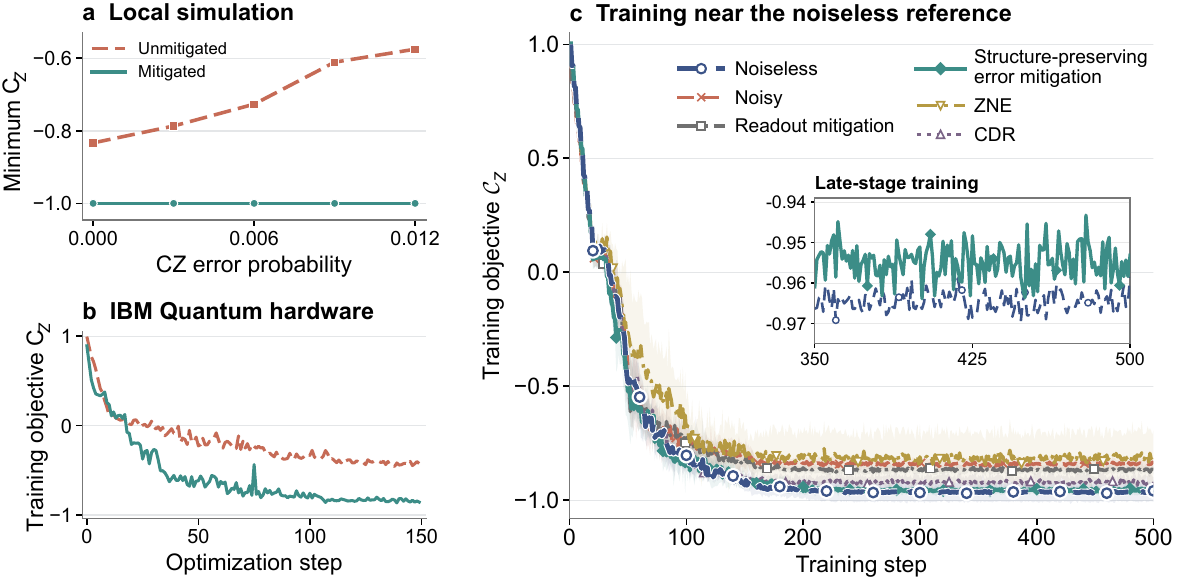}
\caption{Reusable calibration improves variational optimization.
	(a) Minimum magnetization $\mathcal{C}_Z=\langle Z\rangle$
	[Eq.~\eqref{z}] reached during optimization as a function of the
	CZ-gate error probability. Mitigation keeps the optimized
	magnetization close to the ideal minimum of $-1$ across the
	simulated noise range.
	(b) Experimental optimization trajectories on an IBM quantum
	processor. After 150 optimization steps, the mitigated
	magnetization reaches $-0.86$, compared with $-0.42$ without
	mitigation. Teal solid and orange dashed curves denote mitigated
	and unmitigated results, respectively.
	(c) A fixed calibration brings training close to the noiseless reference, yielding the closest final mean training objective among the noisy protocols tested: $\mathcal{C}_Z=-0.953$ versus $-0.958$ in the noiseless control. The dark blue dashed line with open circles denotes noiseless training; the teal solid line with filled diamonds denotes structure-preserving error mitigation. The inset magnifies the unsmoothed mean trajectories of these two methods over steps 350--500. Other curves show unmitigated training, readout mitigation, zero-noise extrapolation (ZNE), and Clifford data regression (CDR). All conditions use the same objective and zero initial angles, with at most 500 objective evaluations per run. Noisy conditions use the CZ-associated error parameter $e_2=0.006$ and a readout-flip probability of $0.01$. Curves show the mean over five independent finite-shot repeats; shaded bands in the main panel indicate one sample standard deviation. Simulations in (a,c) use five qubits and four circuit layers. Protocol details and measurement budgets are given in Appendix~\ref{app:training_comparison}.}
	\label{fig:data6}
\end{figure*}

\subsection{Error mitigation improves variational optimization in simulation and on quantum hardware}

To assess the effectiveness of our error-mitigation strategy, we first isolate the trainable quantum block from the full QNN architecture. Specifically, we remove the classical preprocessing, data-encoding, and postprocessing layers, and optimize only the variational ansatz. This controlled setting provides a direct test of whether mitigation improves the intrinsic trainability of a noisy parameterized circuit. The corresponding noise-response model is developed in Appendix~\ref{app:noise_effects_optimization}.

We consider the normalized average Pauli-$Z$ magnetization for the initial state $\ket{\psi_0}=\ket{0}^{\otimes N}$:
\begin{equation}
    \mathcal{C}_{Z}(\boldsymbol{\theta})
    =
    \frac{1}{N}
    \bra{\psi_{0}}
    V^{\dagger}_{\mathrm{train}}(\boldsymbol{\theta})
    \left(
    \sum_{q=1}^{N} Z_q
    \right)
    V_{\mathrm{train}}(\boldsymbol{\theta})
    \ket{\psi_{0}}.
    \label{z}
\end{equation}
Here, $V_{\mathrm{train}}(\boldsymbol{\theta})$ is the trainable variational ansatz. This objective is particularly relevant because it is built from the same local Pauli-$Z$ expectation values used as quantum features in the full QNN. Thus, the benchmark directly probes the quality of the measured signals that later enter the QNN loss function. Noise affects this optimization by contracting the measured magnetization and reducing its parameter sensitivity. Schematically, we have $\mathcal{C}_{Z}^{\rm noisy}(\boldsymbol{\theta})
    \simeq
    \lambda_Z
    \mathcal{C}_{Z}^{\rm ideal}(\boldsymbol{\theta})$, which is the scalar counterpart of Eq.~\eqref{eq:noisy_feature_simple}. Our mitigation protocol compensates this architecture-dependent contraction by applying the calibrated feature estimator $\eta$, which acts as a regularized inverse of the response on the measured features. In this way, it restores part of the lost observable contrast and improves the effective resolvability of different parameter points during training.

We first evaluate this benchmark using a local noisy simulator, as summarized in Fig.~\ref{fig:data6}(a). For each CZ error probability, we record the minimum objective value reached during training. Without mitigation, the optimized observable deteriorates steadily as the error probability increases, indicating that noise suppresses the useful parameter dependence of the circuit output. In contrast, the mitigated minimum remains near the ideal value of $-1$ over the full range of error probabilities considered. This shows that the structure-preserving calibration compensates the dominant architecture-level distortion of the measured observables and helps maintain the effective optimization landscape. The simulated noise channels are specified in Appendix~\ref{app:qnn_noise}.

We then implement the same benchmark directly on IBM Quantum hardware (Appendix~\ref{app:ibm}, Fig.~\ref{fig:ibm}). Full end-to-end QNN training on hardware is costly because each optimization step requires many finite-shot circuit evaluations, and reliable performance assessment typically requires multiple independent runs. To obtain a practical hardware-level validation, we therefore use the simplified magnetization benchmark in Eq.~\eqref{z} and restrict the optimization to 150 iterations. As shown in Fig.~\ref{fig:data6}(b), the error-mitigated trajectory converges faster and reaches lower cost values than the unmitigated baseline, ending at $\mathcal{C}_Z=-0.86$ compared with $-0.42$. This demonstrates that the proposed mitigation protocol remains effective under realistic noise and can be integrated into iterative, closed-loop variational optimization.

In local training simulations [Fig.~\ref{fig:data6}(c)], structure-preserving error mitigation substantially improves variational optimization, bringing the training objective close to the noiseless reference. After 500 objective evaluations, the mitigated circuit achieves $\mathcal{C}_Z=-0.953$, compared with $-0.958$ for noiseless training and $-0.839$ without mitigation. In comparison, readout mitigation, zero-noise extrapolation (ZNE), and Clifford data regression (CDR) yield $\mathcal{C}_Z=-0.866$, $-0.817$, and $-0.932$, respectively. Among the noisy protocols considered, our method achieves the closest agreement with the noiseless result. The improvement over conventional readout mitigation highlights the importance of accounting for errors accumulated during circuit execution, rather than correcting measurement errors alone. Notably, this improvement is achieved using a single architecture-specific calibration, whose estimator remains fixed throughout optimization despite changes in the variational parameters. These results demonstrate that reusable circuit-level calibration can effectively improve variational optimization without requiring repeated calibration or additional mitigation circuits at each training step.

\begin{figure*}[!t]
	\centering
	\includegraphics[width=0.7\linewidth]{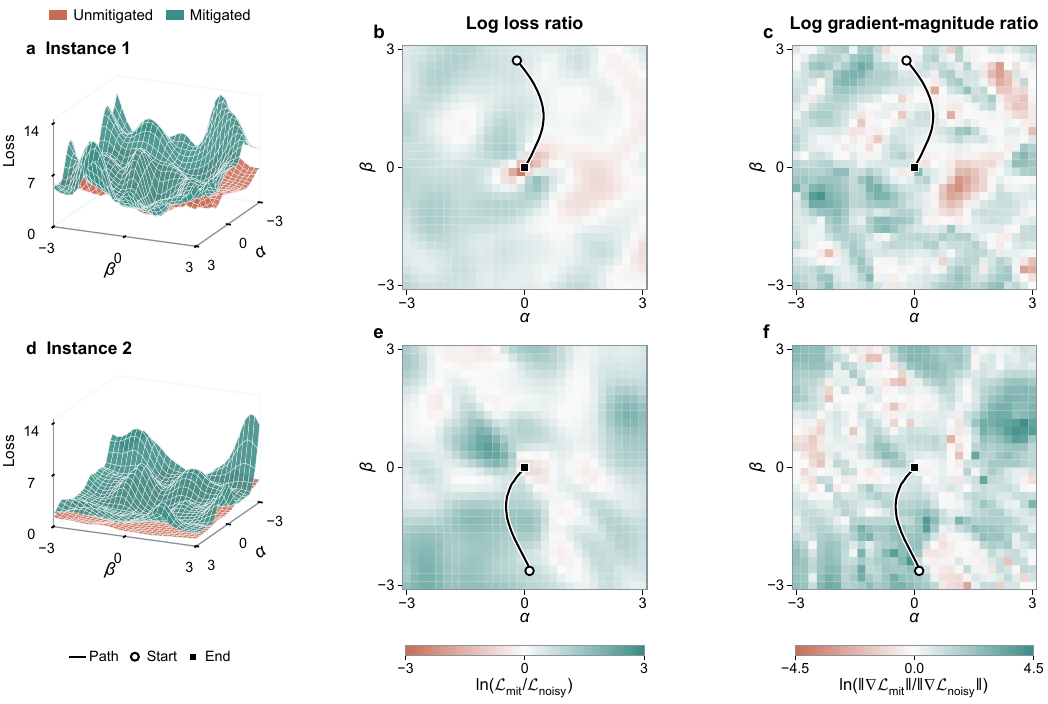}
	\caption{{Calibration increases loss variation and gradient magnitude over much of the displayed PCA plane in two numerical classification examples.}
		The upper row (Instance 1) and lower row (Instance 2) correspond to 10-class Digits classification with two four-qubit channels and binary Breast Cancer classification with one four-qubit channel, respectively. Both are simulated with a two-qubit error parameter $e_2=0.08$.
		(a,d) Mitigated (teal) and unmitigated (orange) loss surfaces, $\mathcal{L}_{\mathrm{mit}}$ and $\mathcal{L}_{\mathrm{noisy}}$, projected onto the two-dimensional PCA plane. The horizontal coordinates $\alpha$ and $\beta$ represent $\alpha_{\mathrm{PCA}}$ and $\beta_{\mathrm{PCA}}$, respectively, while the vertical axis gives the loss $\mathcal{L}$. 
		(b,e) Logarithmic loss ratio, $\ln(\mathcal{L}_{\mathrm{mit}}/\mathcal{L}_{\mathrm{noisy}})$, evaluated at corresponding points in the PCA plane. Negative (red) and positive (teal) values indicate lower and higher mitigated loss, respectively.
		(c,f) Logarithmic gradient-magnitude ratio, $\ln[(G_{\mathrm{mit}}+\epsilon)/(G_{\mathrm{noisy}}+\epsilon)]$, where $G=\|\nabla_{\alpha_{\mathrm{PCA}},\beta_{\mathrm{PCA}}}\mathcal{L}\|_2$ and {$\epsilon=10^{-12}$ is the numerical regularizer.} Positive (teal) values indicate larger gradients under mitigation.
		Dark curves denote the projected optimization trajectories, with open circles and filled squares marking the initial and final parameter coordinates, respectively.}
	\label{fig:loss}
\end{figure*}

\subsection{{Feature calibration and QNN loss contrast}}\label{sec:landscape_revision}

{The above magnetization benchmark shows that a fixed feature correction can preserve the parameter dependence needed for optimization; in Fig.~\ref{fig:data6}(c), it also brings the final mean objective closest to the noiseless reference among the tested noisy protocols. We now ask how correcting these features changes the loss of the full classifier.} For training, the central question is not only whether a measured observable is biased, but also whether parameter-dependent changes in the loss remain resolvable under finite sampling~\cite{kubler2020adaptive,sweke2020stochastic}. {At fixed classical weights, a circuit parameter affects the loss through the measured features~\cite{stokes2020natural}. In the parameter-independent scalar approximation of Eq.~\eqref{eq:noisy_feature_simple}, their Jacobian can be described as follows:}
\begin{equation} \partial_{\boldsymbol{\theta}}\mathbf{s}^{\mathrm{noisy}}
    \simeq
\lambda_{\rm eff}\partial_{\boldsymbol{\theta}}\mathbf{s}^{\mathrm{ideal}} .
    \label{eq:gradient_damping}
\end{equation}
When the reference is correctly labelled and the effective response transfers to the trained circuit, feature calibration can compensate for this architecture-dependent contraction. In a scalar approximation, this gives
\begin{equation}
    \mathbf{s}^{\mathrm{mit}}
    \simeq
    \lambda^{-1}_{\mathrm{mit}}
    \mathbf{s}^{\mathrm{noisy}},
    \label{eq:mit_signal}
\end{equation}
where $\lambda^{-1}_{\mathrm{mit}}$ represents a calibrated inverse contraction. The corresponding mean derivative is schematically $\lambda^{-1}_{\mathrm{mit}}\lambda\partial_{\boldsymbol{\theta}}\mathbf{s}^{\mathrm{ideal}}$, which recovers the ideal derivative when $\lambda_{\mathrm{mit}}=\lambda$. Appendix~\ref{supp:observable_recovery} gives exact feature and derivative errors for a fixed estimator beyond this scalar picture.

\begin{figure*}[!t]
	\centering
	\includegraphics[width=0.7\linewidth]{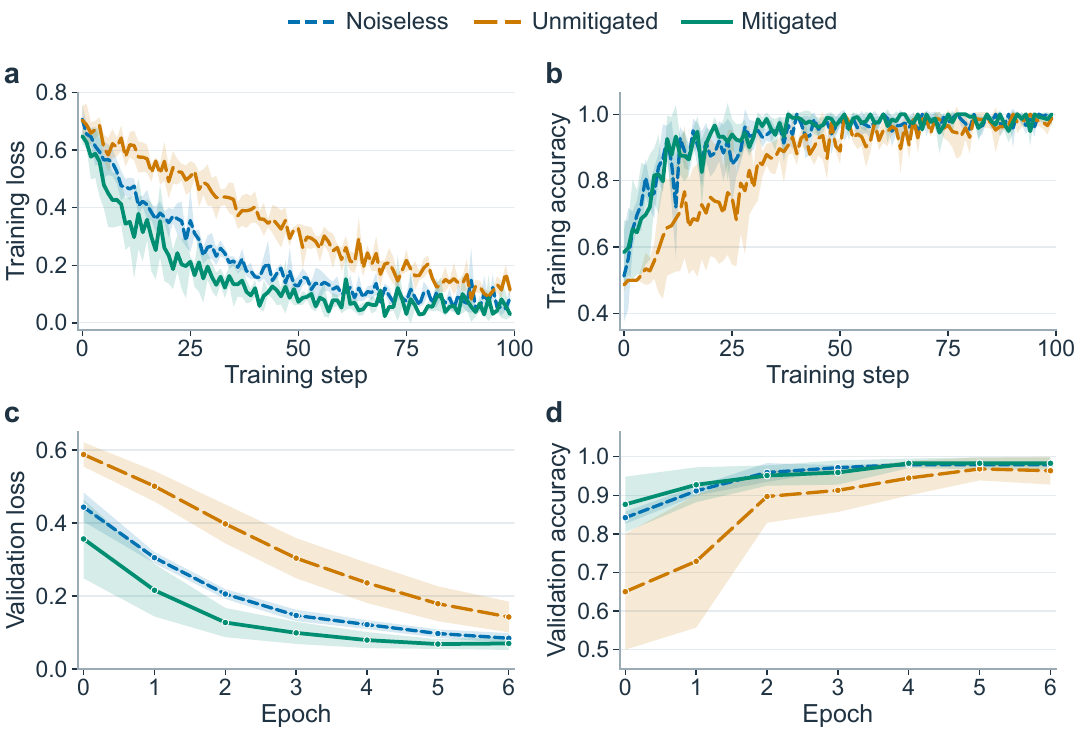}
	\caption{Error mitigation improves QNN training and classification performance.
		(a) Training loss over 100 optimization steps. Mitigation accelerates loss reduction and achieves lower loss than that obtained with either unmitigated or noiseless training.
		(b) Training accuracy, showing that mitigation enables faster convergence towards nearly perfect classification, whereas unmitigated training requires substantially more optimization steps.
		(c) Validation loss over seven epochs. Mitigation consistently yields the lowest validation loss, indicating that the improvement extends beyond the training data.
		(d) Validation accuracy, demonstrating faster convergence and improved classification performance under mitigation, approaching the noiseless reference while outperforming unmitigated training.
		Blue, orange and teal curves denote noiseless, unmitigated and mitigated results, respectively. Curves show the mean over three independent runs, with shaded bands indicating one standard deviation.
		The binary Breast Cancer classifier comprises two four-qubit channels with four circuit repetitions. Local simulations use 2,048 shots per circuit, a single-qubit error rate $e_1=0$, a two-qubit error rate $e_2=0.05$ and a readout error probability of 0.01. Calibration uses 4,000 shots per basis input.}
	\label{fig:noisyqnn}
\end{figure*}

To visualize this effect, we project the variational parameter space onto a two-dimensional principal-component-analysis (PCA) plane, with coordinates $(\alpha_{\mathrm{PCA}},\beta_{\mathrm{PCA}})$ along the first two principal directions~\cite{jolliffe2016principal}. Figure~\ref{fig:loss} presents two representative classification tasks, 10-class Digits and binary Breast Cancer. The mitigated and unmitigated loss surfaces, $\mathcal{L}_{\mathrm{mit}}$ and $\mathcal{L}_{\mathrm{noisy}}$, are evaluated in the same PCA plane, allowing a direct comparison of their optimization landscapes [Figs.~\ref{fig:loss}(a,d)]. In both tasks, the unmitigated surface is comparatively flat, whereas the mitigated surface shows pronounced peaks and valleys over the same parameter range.

To quantify the changes in the loss landscape, we evaluate the logarithmic loss ratio, $\ln(\mathcal{L}_{\mathrm{mit}}/\mathcal{L}_{\mathrm{noisy}})$, at each point in the PCA plane, with the projected training trajectory overlaid [Figs.~\ref{fig:loss}(b,e)]. Negative values indicate regions where mitigation reduces the loss, whereas positive values indicate a larger mitigated loss. The ratio is positive over most of the plane, while negative values are confined to smaller regions, including the vicinity of the trajectory end point.

We further characterize the local parameter sensitivity through the logarithmic gradient-magnitude ratio, $\ln[(G_{\mathrm{mit}}+\epsilon)/(G_{\mathrm{noisy}}+\epsilon)]$, where $G=\|\nabla_{\alpha_{\mathrm{PCA}},\beta_{\mathrm{PCA}}}\mathcal{L}\|_2$ denotes the gradient magnitude within the PCA plane and $\epsilon>0$ is a small numerical regularizer [Figs.~\ref{fig:loss}(c,f)]. Positive values indicate enhanced local gradient magnitude under mitigation. In both tasks, the gradient ratio is positive over most of the plane, with scattered regions of reduced gradient, revealing a broad increase in parameter sensitivity that can influence the optimization dynamics. The distinction between feature and derivative recovery is analyzed in Appendix~\ref{supp:observable_recovery}.

\begin{figure*}[!t]
	\centering
	\includegraphics[width=0.7\linewidth]{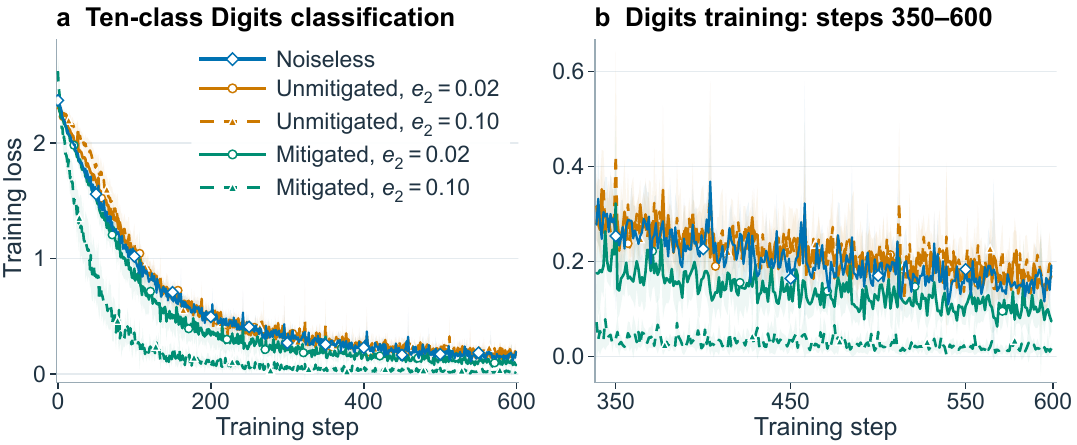}
	\caption{Structure-preserving error mitigation reduces training loss in ten-class Digits classification.
		(a) Training loss over 600 optimization steps. Mitigation accelerates convergence and reduces the training loss at both noise strengths, with a particularly pronounced improvement at the higher two-qubit error rate.
		(b) Training loss at steps 350--600, showing that the loss reduction persists late in training. At $e_2=0.10$, the mitigated loss approaches zero, whereas unmitigated training remains at substantially higher loss.
		Blue curves represent noiseless training, while orange and teal denote unmitigated and mitigated training, respectively. Solid lines with open circles correspond to $e_2=0.02$, and dashed lines with filled triangles to $e_2=0.10$. Curves show the mean over five independent runs, with shaded bands indicating one standard deviation.
		The QNN comprises two four-qubit channels with two circuit repetitions. Local simulations use 1,024 shots per circuit, a single-qubit error rate $e_1=0$, a readout error probability of 0.01, and 10,000 calibration shots per computational-basis input.}
	\label{fig:data2}
\end{figure*}

\subsection{Error mitigation improves QNN training in binary and 10-class tasks}

{For the two instances in Fig.~\ref{fig:loss}, mitigation increases loss variation and gradient magnitude over most of the displayed PCA plane.} We next follow complete training runs to test how this restored signal changes the training dynamics of noisy QNNs \cite{mcclean2018barren,cerezo2021cost,wang2021noise,larocca2024review,grant2019initialization,skolik2021layerwise,pesah2021absence,li2023ensemble}. Because hardware-based training requires repeated circuit evaluations for loss and gradient estimation, its measurement cost becomes substantial, particularly across multiple initializations and training runs \cite{preskill2018quantum,schuld2019evaluating,arrasmith2021effect,bharti2022noisy,huggins2021efficient,gonthier2022measurements,wierichs2022shift,mari2021higher,shen2026portfolio}. Unless otherwise stated, we therefore use a local noisy simulator to systematically compare noiseless, noisy, and error-mitigated training under controlled noise and sampling conditions \cite{czarnik2021error,lowe2021unified,shen2025structure,ahmed2025comparative,ahmed2025noisy}. The noise model is specified in Appendix~\ref{app:qnn_noise}.

We first benchmark binary classification on the Breast Cancer dataset. Dataset splitting and preprocessing are described in Appendix~\ref{app:qnn_data}. Figures~\ref{fig:noisyqnn}(a,b) show the training loss and training accuracy over 100 optimization steps. In the noiseless case, the QNN rapidly approaches a low-loss region and achieves high training accuracy. In contrast, under noisy execution, optimization becomes slower: the loss decreases more gradually and saturates at a higher value, consistent with the noise-induced suppression of gradients discussed above. Applying structure-preserving error mitigation largely restores the learning dynamics. The mitigated trajectory follows the noiseless reference more closely, reaches lower loss values, and converges faster than the unmitigated noisy baseline.

We also evaluate generalization performance during training. Figures~\ref{fig:noisyqnn}(c,d) show the corresponding validation loss and validation accuracy, with shaded regions indicating the standard deviation over independent runs. Without mitigation, the validation curves exhibit stronger fluctuations and systematically degraded accuracy. By contrast, the mitigated QNN achieves higher mean validation accuracy and lower validation loss, while also reducing run-to-run variability. These results indicate that the mitigation procedure improves not only the training trajectory but also the stability of the learned classifier.

We next consider the more demanding 10-class Digits image classification task and test the robustness of the method under stronger two-qubit gate noise. In this setting, we vary the per-endpoint bit-flip parameter $e_2$ attached to those gates. The circuit implementation and local noise model are given in Appendices~\ref{app:qnn_circuit} and~\ref{app:qnn_noise}. Figure~\ref{fig:data2}(a) shows the training loss for noiseless circuits and for noisy circuits with two-qubit error rates $e_2=0.02$ and $e_2=0.10$, with and without mitigation. Consistent with the above binary benchmark, unmitigated noise slows convergence and leads to higher training loss, whereas our error-mitigation method substantially improves the training trajectory, producing faster convergence and lower final loss across the noise strengths considered. A closer view of the late-stage training regime is shown in Fig.~\ref{fig:data2}(b), focusing on optimization steps 350--600. The mitigated trajectories reach lower loss plateaus and display reduced oscillations compared with the unmitigated noisy cases.

\begin{figure*}[!t]
    \centering
    \includegraphics[width=0.99\linewidth]{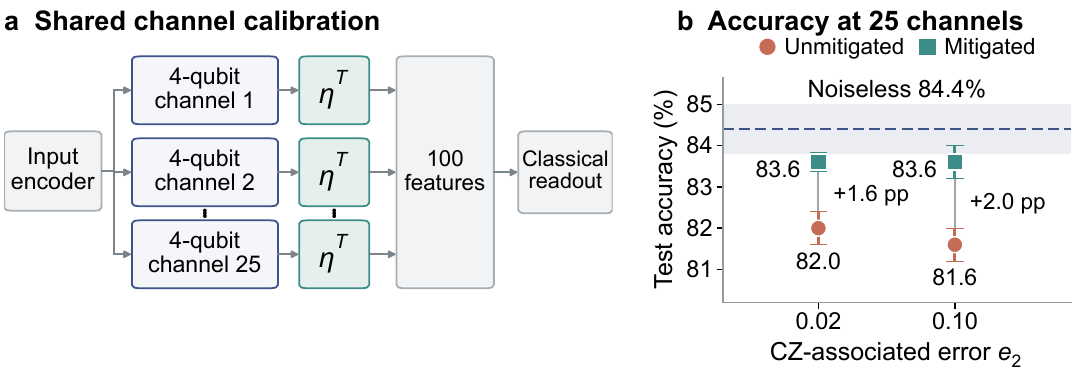}
    \caption{Reusable channel calibration in a 25-channel QNN with 100 aggregate qubits.
    (a) The quantum layer comprises 25 independent four-qubit channels, whose 100 local Pauli-$Z$ features enter a classical readout. There are no inter-channel quantum gates. Channels with compatible effective responses share a $16\times16$ response matrix and a $16\times4$ feature estimator $\eta$.
    (b) Reported mean test accuracies at $e_2=0.02$ and $0.10$. Structure-preserving error mitigation raises accuracy from 82.0\% to 83.6\% and from 81.6\% to 83.6\%, respectively. The remaining gap is 0.8 percentage points. Terracotta circles and teal squares show unmitigated and mitigated means; the navy dashed line denotes the noiseless reference. Error bars and shading show the retained reported standard deviations, not confidence intervals.}
    \label{fig:qnncompare}
\end{figure*}

\subsection{Calibration reuse in a 25-channel modular QNN}

We next consider calibration reuse as the number of channels increases. The architecture in Fig.~\ref{fig:qnncompare}(a) comprises 25 independent four-qubit channels: its 100 qubits are counted in aggregate, with quantum gates acting within each four-qubit block and a classical readout combining the channel features. For a modular architecture comprising $P$ channels of $N$ qubits each, the quantum layer generates $PN$ measurement features, while calibration requires a $2^N\times2^N$ response matrix and a $2^N\times N$ feature estimator per channel. When channels share a compatible effective noise response, one calibrated estimator can be reused throughout the network. The estimator construction and requirements for reuse are detailed in Appendices~\ref{sec:mitigation} and~\ref{supp:observable_recovery}.

Figure~\ref{fig:qnncompare}(b) shows the retained accuracy comparison for the 25-channel model. At $e_2=0.02$ and $0.10$, the unmitigated mean test accuracies are 82.0\% and 81.6\%, respectively; calibration increases both to 83.6\%. Although modest in absolute terms, these gains of 1.6 and 2.0 percentage points recover approximately 67\% and 71\% of the respective 2.4- and 2.8-percentage-point gaps to the noiseless mean of 84.4\%. The residual gap is 0.8 percentage points at both noise strengths. This comparison quantifies recovery of noise-degraded accuracy within the same architecture.

For four-qubit channels, a shared calibration uses 16 computational-basis input settings and one $16\times16$ response matrix, independent of $P$, provided that the effective responses remain compatible. Distinct channel responses instead require $16P$ settings and $P$ small response matrices. At $P=25$, these alternatives use 16 and 400 settings, respectively, rather than the $2^{100}$ settings of a dense full-system basis calibration. Channel execution and feature-processing costs still grow with $P$, and the shot cost equals the number of calibration settings multiplied by the shots per setting. Together with the training results in Figs.~\ref{fig:noisyqnn} and~\ref{fig:data2}, the 25-channel accuracy comparison illustrates how local calibration can be reused in a wider modular QNN without enlarging the calibrated block.

\section{Discussion}\label{discussion}

{Quantum noise can degrade QNN trainability by distorting the measured quantum features and their dependence on variational parameters, thereby modifying the loss landscape and the gradients available to the optimizer \cite{mcclean2018barren,cerezo2021cost,wang2021noise,arrasmith2021effect,larocca2024review}. Our approach addresses this problem by calibrating the effective noise response of a fixed circuit architecture and directly correcting the observables that determine the learning objective. Rather than characterizing individual microscopic noise processes, the method exploits the shared circuit structure to construct a reusable feature estimator \cite{temme2017error,endo2018practical,cai2023quantum,shen2025structure}. Crucially, the theoretical analysis establishes that successful reuse requires not only accurate calibration of the reference circuit but also sufficiently small response mismatch and parameter-dependent variations across the training circuit family. Under these conditions, the calibrated estimator can recover both the measured features and their parameter sensitivities, providing a direct mechanism for improving optimization under noise.}

{Our numerical results demonstrate the consequences of this feature-level correction for variational learning. Across the classification benchmarks, structure-preserving mitigation increases loss contrast and gradient magnitudes over substantial regions of the parameter space, accelerates convergence, and improves training and validation performance relative to unmitigated noisy execution. These improvements indicate that recovering the parameter dependence of measured observables can substantially improve the effective optimization landscape, beyond correcting biases in individual expectation values. The IBM Quantum experiment further demonstrates that the same calibration strategy improves closed-loop variational optimization on superconducting quantum hardware, supporting its practical applicability beyond idealized noise models. These findings complement existing approaches that integrate error mitigation into variational optimization \cite{kreplin2024reduction,robbiati2026realtime}.}

An important practical advantage is that the calibration can be reused across training iterations and, when the effective noise responses are compatible, across independent quantum channels. For a modular QNN comprising $P$ channels of $N$ qubits, a shared calibration requires a $2^N\times 2^N$ response matrix, independent of the total number of channels. This reduces the calibration requirements relative to a full-system treatment, although circuit execution and feature processing still scale with $P$. The benefit therefore lies in the reuse of a small, architecture-specific calibration rather than the elimination of sampling costs or an unconditional guarantee of transfer across circuits~\cite{takagi2023universal,tsubouchi2023universal,quek2024limitations}. In this respect, our method complements readout mitigation, learned corrections, and other circuit- or structure-aware mitigation strategies \cite{bravyi2021mitigating,nation2021scalable,czarnik2021error,lowe2021unified,wang2024can,korolev2026machine,tran2023locality,eddins2024lightcone,filippov2023tensor}.

\section*{ACKNOWLEDGMENTS}\label{sec}
We acknowledge the use of IBM Quantum services for this work. The views expressed are those of the authors and do not reflect the official policy or position of IBM or the IBM Quantum team.
This research is supported by the Singapore Ministry of Education (MOE) through an Academic Research Fund Tier-II Grant (MOE-T2EP50224-0007). Additional support was provided by the {Air Force Office of Scientific Research} (Award No. FA2386-24-1-4011).

\bibliographystyle{apsrev4-2}
\bibliography{ref}

\clearpage
\appendix
\onecolumngrid
\counterwithout{equation}{section}
\setcounter{equation}{0}
\setcounter{figure}{0}
\renewcommand{\theequation}{S\arabic{equation}}
\renewcommand{\theHequation}{S\arabic{equation}}
\renewcommand{\thetable}{S\arabic{table}}
\renewcommand{\thefigure}{S\arabic{figure}}
\renewcommand{\theHfigure}{S\arabic{figure}}
\renewcommand{\thesection}{S\arabic{section}}
\makeatletter
\renewcommand{\p@subsection}{\thesection.}
\makeatother
\setcounter{page}{1}
\renewcommand{\thepage}{S\arabic{page}}

% \begin{center}
% {\huge Supplementary Information}
% \end{center}

\section{Notation and conventions}

{This glossary connects the circuit and estimator labels in Fig.~\ref{fig:fig1} to the quantities used in the main-text equations and the recovery analysis below. In particular, it distinguishes channel size, aggregate width and the several unrelated uses of contraction and regularization parameters.}

\label{sec:glossary}

All probability, feature, logit, and label vectors are column vectors. Sample
and channel indices are omitted when the context refers to one sample or one
channel. In a single-channel formula, $\mathbf{s}\in\mathbb{R}^{N}$ and
$\mathbf{p}\in\mathbb{R}^{2^N}$. In a formula for the complete QNN,
$\mathbf{s}\in\mathbb{R}^{NP}$ is the column obtained by stacking channel features.
The superscripts $\mathrm{ideal}$, $\mathrm{noisy}$, and $\mathrm{mit}$ identify
the execution or estimation condition, and $(\alpha)$ identifies a quantum channel.
Shortened figure labels and schematic aliases are listed with the corresponding
symbols below.

\textbf{Indices}
\begin{itemize}[leftmargin=3.2cm,labelsep=0.6em,itemsep=0pt,topsep=0pt]
    \item[$i=1,\ldots,Q$] Sample index.
    \item[$k,k'=1,\ldots,C$] True, predicted, or summation class indices.
    \item[$\alpha=1,\ldots,P$] {Quantum-channel index; Fig.~\ref{fig:fig1}(c) displays channels $1,2,\ldots,P$.}
    \item[$q,q'=1,\ldots,N$] Qubit indices within a channel.
    \item[$\ell$] Variational-layer index, with $L$ counting entangling repetitions. The CX ansatz used in the classification benchmarks has rotation layers $1,\ldots,L+1$, while the CZ schematic has $L$. Distinct from the logit vector $\mathbf{l}$.
    \item[$b,b'$] Computational-basis bitstring indices in $\{0,\ldots,2^N-1\}$. In calibration, $b$ is the prepared input and $b'$ the measured output.
    \item[$j$] Coordinate index for the quantum-parameter derivative $\partial_j=\partial/\partial\theta_j$.
\end{itemize}

\vspace{0.5em}
\textbf{Problem sizes and shot counts}
\begin{itemize}[leftmargin=3.2cm,labelsep=0.6em,itemsep=0pt,topsep=0pt]
    \item[$Q,C,K$] Dataset sample count, class count, and original input dimension, respectively.
    \item[$P,N,L$] Number of parallel quantum channels, qubits per channel, and entangling repetitions, respectively. The aggregate number of channel qubits is $NP$.
    \item[$S_{\mathrm{calib}}$] Calibration shots per prepared basis state.
\end{itemize}

\vspace{0.5em}
\textbf{Data and classical preprocessing/postprocessing}
\begin{itemize}[leftmargin=3.2cm,labelsep=0.6em,itemsep=0pt,topsep=0pt]
    \item[$\mathbf{x}_i$] Input for sample $i$, in $\mathbb{R}^{K}$.
    \item[$\mathbf{x}_{\mathrm{latent},i}$] Latent column in $\mathbb{R}^{NP}$, equal to $W_{\mathrm{pre}}\mathbf{x}_i+\mathbf{b}_{\mathrm{pre}}$.
    \item[$\mathbf{x}^{(\alpha)}_i$] Length-$N$ latent subvector supplied to channel $\alpha$.
    \item[$\mathbf{y}_i$] One-hot label in $\{0,1\}^{C}$, with components $y_{i,k}$.
    \item[$\mathbf{l}_i,\mathbf{f}_i$] Length-$C$ logit and predicted-probability columns, with components $l_{i,k}$ and $f_{i,k}$. Softmax maps $\mathbf{l}_i$ to $\mathbf{f}_i$.
    \item[$W_{\mathrm{pre}},\mathbf{b}_{\mathrm{pre}}$] Preprocessing weights and bias, of dimensions $NP\times K$ and $NP$.
    \item[$W_{\mathrm{post}},\mathbf{b}_{\mathrm{post}}$] Readout weights and bias, of dimensions $C\times NP$ and $C$, with $\mathbf{l}=W_{\mathrm{post}}\mathbf{s}+\mathbf{b}_{\mathrm{post}}$.
    \item[$\mathcal{W}$] Collection $\{W_{\mathrm{pre}},\mathbf{b}_{\mathrm{pre}},W_{\mathrm{post}},\mathbf{b}_{\mathrm{post}}\}$ of classical parameters, denoted $w$ in Fig.~\ref{fig:fig1}.
\end{itemize}

\vspace{0.5em}
\textbf{Quantum circuits, states, and measurements}
\begin{itemize}[leftmargin=3.2cm,labelsep=0.6em,itemsep=0pt,topsep=0pt]
    \item[$V$] Encoded channel circuit $V(\mathbf{x}^{(\alpha)}_i,\boldsymbol{\theta}^{(\alpha)})=V_{\mathrm{train}}(\boldsymbol{\theta}^{(\alpha)})V_{\mathrm{enc}}(\mathbf{x}^{(\alpha)}_i)$. {Figure~\ref{fig:fig1}(a) labels this channel block $V$, while panel (c) uses $U_{\theta,\alpha}$ for channel $\alpha$.}
    \item[$V_{\mathrm{enc}},V_{\mathrm{train}}$] Data-encoding unitary and trainable variational ansatz.
    \item[$V_{\mathrm{rot}}^{(\ell)},V_{\mathrm{ent}}$] Layer-$\ell$ rotations and the fixed entangling pattern, specified separately for the CX implementation and CZ example. Operator products act on states from right to left.
    \item[$\boldsymbol{\theta}$] Quantum parameters, with $\boldsymbol{\theta}^{(\alpha)}$ for a channel and $\theta_{q,y}^{(\ell)},\theta_{q,z}^{(\ell)}$ for its layer rotations.
    \item[$\ket{\psi_0},\rho_0$] Reference state $\ket{0}^{\otimes N}$ and projector $\rho_0=\ket{\psi_0}\bra{\psi_0}$.
    \item[$\ket{\psi}$] Encoded channel output state $\ket{\psi(\mathbf{x}^{(\alpha)},\boldsymbol{\theta}^{(\alpha)})}=V\ket{\psi_0}$.
    \item[$X_q,Y_q,Z_q$] Local Pauli operators on qubit $q$.
    \item[$s_q^{(\alpha)},\mathbf{s}^{(\alpha)}$] Scalar local expectation $\langle Z_q^{(\alpha)}\rangle$ [Eq.~\eqref{si}] and its length-$N$ feature column.
    \item[$s,\Delta s$] One measured feature $s_q$, drawn as $s(\theta)$ against one circuit parameter in Fig.~\ref{fig:fig1}(d), and its change $\Delta s=s(\theta)-s(\theta')$ between two parameter values.
    \item[$\mathbf{s}$] Length-$NP$ column formed by stacking $\mathbf{s}^{(1)},\ldots,\mathbf{s}^{(P)}$ when all channels are included.
    \item[$\mathbf{p}^{\mathrm{ideal}},\mathbf{p}^{\mathrm{noisy}}$] Length-$2^N$ single-channel probability columns, written $[\mathbf{p}^{(\alpha)}]^{\mathrm{noisy}}$ for channel $\alpha$. The implemented estimator computes corrected features directly, without requiring a reconstructed probability column.
    \item[$F$] Measurement-eigenvalue matrix in $\{-1,+1\}^{2^N\times N}$, with $F_{bq}=\langle b|Z_q|b\rangle$ and $\mathbf{s}^{\mathrm{ideal}}=F^T\mathbf{p}^{\mathrm{ideal}}$.
    \item[$\mathcal{C}_Z$] Normalized mean magnetization $N^{-1}\sum_q\langle Z_q\rangle$ [Eq.~\eqref{z}], plotted as $\langle Z\rangle$ in Fig.~\ref{fig:data6}.
\end{itemize}

\vspace{0.5em}
\textbf{Noise parameters and phenomenological responses}
\begin{itemize}[leftmargin=3.2cm,labelsep=0.6em,itemsep=0pt,topsep=0pt]
    \item[$e_1,e_2,e_{\mathrm{ro}}$] Single-qubit depolarizing parameter, independent per-endpoint Pauli-$X$ error probability after each CX or CZ gate of the implemented circuit, and readout bit-flip probability, respectively. Figure~\ref{fig:data2} labels $e_2$ as $p_2$.
    \item[$p_a$] Binary Pauli-error weights in Eq.~\eqref{eq:two_qubit_noise}, $p_0=1-e_2$ and $p_1=e_2$, distinct from the probability vector $\mathbf{p}$.
    \item[$\lambda,\lambda_Z,\lambda_{\mathrm{mit}}$] Phenomenological feature contraction, mean-magnetization contraction, and calibrated contraction, respectively, with $\lambda_{\mathrm{mit}}^{-1}$ the inverse signal factor. These differ from the regularization strength $\lambda_{\mathrm{reg}}$.
    \item[$A_{\mathrm{noise}},A_{\mathrm{mit}}^{-1}$] Effective $N\times N$ feature response and correction in the single-channel phenomenological model of Sec.~\ref{app:noise_effects_optimization}.
    \item[$\xi_Z,\boldsymbol{\xi}$] Scalar magnetization and vector feature fluctuations.
\end{itemize}

\vspace{0.5em}
\textbf{Error mitigation}
\begin{itemize}[leftmargin=3.2cm,labelsep=0.6em,itemsep=0pt,topsep=0pt]
    \item[$V_{\mathrm{mit}},\boldsymbol{\theta}_{\mathrm{id}}$] Identity-target reference circuit and its fitted parameters.
    \item[$M_{\mathrm{ref}},\Pi$] Raw reference response $(M_{\mathrm{ref}})_{b'b}=\Pr(b'\mid b)$ and permutation matrix of the ideal reference action [Eqs.~\eqref{eq:mch} and \eqref{eq:reference_relabel}].
    \item[$M_{\mathrm{mit}}$] Effective response $M_{\mathrm{ref}}\Pi^T$ in ideal-output labels, of size $2^N\times2^N$. The Fig.~\ref{fig:fig1} schematic shows the identity-reference case, in which $M_{\mathrm{ch}}\equiv M_{\mathrm{mit}}$. Section~\ref{supp:observable_recovery} abbreviates the effective response as $M$.
    \item[$\eta,\eta_q$] $2^N\times N$ feature-estimator matrix and its column $q$, with $M_{\mathrm{mit}}^T\eta\simeq F$ and $\mathbf{s}^{\mathrm{mit}}=\eta^T\mathbf{p}^{\mathrm{noisy}}$ for one channel.
    \item[$\lambda_{\mathrm{reg}}$] Tikhonov regularization strength. The identity $I$ in the estimator solution is $2^N$-dimensional.
    \item[$\mathbf{s}^{\mathrm{mit}}$] Corrected feature column. For channel $\alpha$, it is written $[\mathbf{s}^{(\alpha)}]^{\mathrm{mit}}\in\mathbb{R}^{N}$, and stacking these columns gives the complete length-$NP$ vector.
\end{itemize}

\vspace{0.5em}
\textbf{Training, landscape coordinates, and supplementary theory}
\begin{itemize}[leftmargin=3.2cm,labelsep=0.6em,itemsep=0pt,topsep=0pt]
    \item[$\boldsymbol{v}$] Complete trainable parameter collection, $\mathcal{W}$ together with $\boldsymbol{\theta}$.
    \item[$\mathcal{L}$] Cross-entropy loss, with single-sample form in Eq.~\eqref{eq:loss} and dataset average in Eq.~\eqref{eq:loss_dataset}. Logarithms are natural.
    \item[$\alpha_{\mathrm{PCA}},\beta_{\mathrm{PCA}}$] PCA-plane coordinates, shortened to $\alpha,\beta$ on the Fig.~\ref{fig:loss} axes and distinct from the channel index $\alpha$.
    \item[$G,\epsilon$] Scalar gradient norm $G=\|\nabla_{\alpha_{\mathrm{PCA}},\beta_{\mathrm{PCA}}}\mathcal{L}\|_2$ and stabilizer $\epsilon=10^{-12}$ in its plotted log ratio.
    \item[$\mathcal{U}_{\boldsymbol{\theta}},\mathcal{C}_{\boldsymbol{\theta}}$] Ideal and noisy quantum channels in Sec.~\ref{supp:observable_recovery}. The star denotes their adjoint action.
    \item[$B_q$] Diagonal corrected measurement observable $\sum_b\eta_{bq}|b\rangle\langle b|$.
    \item[$\mathbf{r},\mathbf{R}_q,e_q$] Transfer residual, calibration-residual column, and scalar feature bias [Eqs.~\eqref{eq:supp_recovery_residuals} and \eqref{eq:supp_recovery_bias}].
    \item[$U,\sigma,\mathcal{N}(q)$] Illustrative ideal unitary, a Pauli string, and the neighbor set of qubit $q$ in the two-layer example. Without a qubit argument, $\mathcal{N}$ denotes the noise channel of Sec.~\ref{app:noise_effects_optimization}.
    \item[$d_{\ell q},\gamma_\ell(\sigma)$] Onsite and entangling-error Pauli contractions in layer $\ell$.
    \item[$K_{xq},K_{yq},d_{1\star q}$] Neighbor-supported Pauli strings and the product of onsite contractions over $q$ and its neighbors [Eq.~\eqref{eq:supp_recovery_star}].
    \item[$a_q,h_q,\mu_q$] Longitudinal, transverse, and identity-reference parity responses, with $\mu_q=a_q$ in the stated two-layer zero-angle reference.
    \item[$\vartheta,m_{Z,q},m_{K,q}$] Angle $\vartheta=\theta_{q,y}^{(2)}$ and the two ideal correlators used in Eq.~\eqref{eq:supp_recovery_two_components}.
    \item[$k_q,\Delta_q$] Regularized scalar score $k_q=a_q/(a_q^2+\lambda_{\mathrm{reg}})$ and adjoint-observable residual in the two-layer example.
\end{itemize}

\section{Observable recovery and calibration reuse in QNN training}
\label{supp:observable_recovery}

The structure-preserving calibration targets the measurement functionals that enter the QNN. Its reuse during training is governed by how faithfully a fixed estimator recovers these functionals as circuit parameters change. Here we express this mechanism at the observable level, derive separate contributions from calibration and transfer errors, and illustrate their effect on features and derivatives in a two-layer circuit. {The observable criterion below gives the feature and derivative recovery conditions underlying the schematic feature correction in Fig.~\ref{fig:fig1}(d); the residual identities also guide interpretation of the numerical loss and gradient maps in Fig.~\ref{fig:loss}.}

\subsection{The calibrated estimator as a measurement observable}

Consider one $N$-qubit channel. Write $M\equiv M_{\mathrm{mit}}$ for its response in ideal-output labels [Eq.~\eqref{eq:reference_relabel}], $F_{bq}=\langle b|Z_q|b\rangle$ for the ideal measurement functional, and $\eta_q$ for column $q$ of the estimator
\begin{equation}
    \eta=(MM^{T}+\lambda_{\mathrm{reg}} I)^{-1}MF,
    \qquad
    \mathbf{s}^{\mathrm{mit}}=\eta^{T}\mathbf{p}^{\mathrm{noisy}}.
    \label{eq:supp_recovery_estimator}
\end{equation}
The regularization parameter $\lambda_{\mathrm{reg}}>0$ balances agreement with the calibration constraints against the magnitude of the estimator weights. For a fixed calibration, all quantities below refer to mean features conditioned on that calibration and do not include fluctuations from output sampling.

Let $\mathcal{U}_{\boldsymbol{\theta}}$ and $\mathcal{C}_{\boldsymbol{\theta}}$ denote the ideal and noisy circuit channels, respectively, and let $\rho$ be the input state. The dependence on the classical input is suppressed in this notation. Computational-basis readout errors can be included in the terminal action of $\mathcal{C}_{\boldsymbol{\theta}}$. Define the diagonal observable $B_q=\sum_b\eta_{bq}|b\rangle\langle b|$. The ideal and corrected features are
\begin{equation}
    s_q^{\mathrm{ideal}}=
    \operatorname{Tr}\!\left[\rho\,\mathcal{U}_{\boldsymbol{\theta}}^{*}(Z_q)\right],
    \qquad
    s_q^{\mathrm{mit}}=
    \operatorname{Tr}\!\left[\rho\,\mathcal{C}_{\boldsymbol{\theta}}^{*}(B_q)\right],
    \label{eq:supp_recovery_adjoint}
\end{equation}
where the star denotes the adjoint channel. Thus, recovery for every input state is equivalent to
\begin{equation}
    \mathcal{C}_{\boldsymbol{\theta}}^{*}(B_q)
    =\mathcal{U}_{\boldsymbol{\theta}}^{*}(Z_q).
    \label{eq:supp_recovery_condition}
\end{equation}
The operator norm of the difference between these two Hermitian operators is exactly the largest absolute feature bias over input states. This criterion concerns the measured observable and does not require reconstruction of the output quantum state.

For an ideal identity reference, basis-input calibration determines the diagonal matrix elements of this relation through $M^{T}\eta_q\simeq F_q$. If both adjoint observables are diagonal, equality of these diagonal entries certifies recovery at the reference. Otherwise, basis-input constraints do not determine the off-diagonal entries. Transfer of this recovery across the parameter family additionally depends on the response of the intervening rotations and noisy gates. Even angle-independent errors at individual gates can produce an angle-dependent effective response after propagation through the rest of the circuit.

\subsection{Feature and derivative errors under a fixed calibration}

For arbitrary circuit inputs and parameters, define the transfer residual $\mathbf{r}$ and the calibration residual $\mathbf{R}_q$ by
\begin{equation}
    \mathbf{r}=\mathbf{p}^{\mathrm{noisy}}-M\mathbf{p}^{\mathrm{ideal}},
    \qquad
    \mathbf{R}_q=M^{T}\eta_q-F_q.
    \label{eq:supp_recovery_residuals}
\end{equation}
This definition leaves all mismatch between the calibrated reference and a deployed circuit in $\mathbf{r}$. In particular, it accommodates parameter dependence, input dependence, imperfect identity fitting, and noise drift. The resulting feature bias $e_q=s_q^{\mathrm{mit}}-s_q^{\mathrm{ideal}}$ satisfies the exact identities
\begin{align}
    e_q&=\mathbf{R}_q^{T}\mathbf{p}^{\mathrm{ideal}}+\eta_q^{T}\mathbf{r},
    \label{eq:supp_recovery_bias}\\
    \partial_j e_q&=\mathbf{R}_q^{T}\partial_j\mathbf{p}^{\mathrm{ideal}}
    +\eta_q^{T}\partial_j\mathbf{r},
    \label{eq:supp_recovery_derivative}
\end{align}
where $\partial_j=\partial/\partial\theta_j$ and $M$ and $\eta$ are held fixed. Since $\|\mathbf{p}^{\mathrm{ideal}}\|_1=1$, these imply
\begin{align}
    |e_q|&\leq\|\mathbf{R}_q\|_{\infty}
    +\|\eta_q\|_{\infty}\|\mathbf{r}\|_1,
    \label{eq:supp_recovery_bias_bound}\\
    |\partial_j e_q|&\leq\|\mathbf{R}_q\|_{\infty}
    \|\partial_j\mathbf{p}^{\mathrm{ideal}}\|_1
    +\|\eta_q\|_{\infty}\|\partial_j\mathbf{r}\|_1.
    \label{eq:supp_recovery_derivative_bound}
\end{align}
Consequently, a reusable calibration recovers features when the calibration residual and the amplified transfer residual are small. Recovery of their derivatives requires control of the variation of that transfer residual as well. Regularization limits amplification through the estimator weights while allowing a finite calibration residual. These identities apply to any fixed estimator and response matrix, independently of a particular gate decomposition or scalar noise approximation. The residuals thus provide measures of feature-recovery accuracy beyond the training loss.

\subsection{Analytical example: longitudinal and transverse attenuation}

For an identity reference composed of Clifford gates with stochastic Pauli errors, the exact probability response takes the form $M_{b'b}=c(b'\oplus b)$, where $c$ is the distribution of effective bit flips and $\oplus$ denotes bitwise addition modulo two. The local parity $F_q$ is an eigenvector: $MF_q=\mu_qF_q$. Since this response matrix is symmetric, Eq.~\eqref{eq:supp_recovery_estimator} becomes
\begin{equation}
    \eta_q=\frac{\mu_q}{\mu_q^2+\lambda_{\mathrm{reg}}}F_q.
    \label{eq:supp_recovery_parity}
\end{equation}
In this regime, calibration corrects the attenuation of each local feature. This reduction concerns the exact reference response. A finite-sample response matrix can also mix different parities.

To examine transfer explicitly, consider the ideal two-layer circuit $U=V_{\mathrm{ent}}V_{\mathrm{rot}}^{(2)}V_{\mathrm{ent}}V_{\mathrm{rot}}^{(1)}$, where $V_{\mathrm{ent}}$ is a product of CZ gates on a fixed simple graph and
$V_{\mathrm{rot}}^{(\ell)}=\bigotimes_q R_z(\theta_{q,z}^{(\ell)})R_y(\theta_{q,y}^{(\ell)})$.
Assume stationary, angle-independent stochastic Pauli errors after the entangling gates and onsite depolarization after each rotation layer. Let $d_{\ell q}>0$ be the onsite Pauli contraction. Propagating the entangling-gate errors through the remaining CZ gates of that layer gives a Pauli channel with eigenvalues $\gamma_\ell(\sigma)>0$ on Pauli strings $\sigma$. For the neighbor set $\mathcal{N}(q)$, introduce
\begin{equation}
    K_{xq}=X_q\prod_{q'\in\mathcal{N}(q)}Z_{q'},
    \qquad
    d_{1\star q}=\prod_{q'\in\{q\}\cup\mathcal{N}(q)}d_{1q'}.
    \label{eq:supp_recovery_star}
\end{equation}
The longitudinal and transverse responses are, respectively,
\begin{equation}
    a_q=\gamma_2(Z_q)d_{2q}\gamma_1(Z_q)d_{1q},
    \qquad
    h_q=\gamma_2(Z_q)d_{2q}\gamma_1(X_q)d_{1\star q}.
    \label{eq:supp_recovery_two_responses}
\end{equation}
Writing $\vartheta=\theta_{q,y}^{(2)}$, $m_{Z,q}=\operatorname{Tr}(\rho [V_{\mathrm{rot}}^{(1)}]^\dagger Z_qV_{\mathrm{rot}}^{(1)})$, and
$m_{K,q}=\operatorname{Tr}(\rho [V_{\mathrm{rot}}^{(1)}]^\dagger K_{xq}V_{\mathrm{rot}}^{(1)})$ gives
\begin{equation}
    s_q^{\mathrm{ideal}}=\cos\vartheta\,m_{Z,q}-\sin\vartheta\,m_{K,q},
    \qquad
    s_q^{\mathrm{noisy}}=a_q\cos\vartheta\,m_{Z,q}-h_q\sin\vartheta\,m_{K,q}.
    \label{eq:supp_recovery_two_components}
\end{equation}
These expressions follow by propagating $Z_q$ backwards: the final CZ layer fixes $Z_q$, the second $R_y$ rotation mixes $Z_q$ and $X_q$, and the first CZ layer transforms $X_q$ into $K_{xq}$. The two resulting components acquire the distinct factors in Eq.~\eqref{eq:supp_recovery_two_responses}.

With all angles zero, $V_{\mathrm{ent}}^2=I$ and the reference measures $\mu_q=a_q$. In the unregularized limit, the score $Z_q/a_q$ therefore leaves the operator residual
\begin{equation}
    \Delta_q\equiv
    \mathcal{C}_{\boldsymbol{\theta}}^*(Z_q/a_q)
    -\mathcal{U}_{\boldsymbol{\theta}}^*(Z_q)
    =\left(1-\frac{h_q}{a_q}\right)\sin\vartheta\,[V_{\mathrm{rot}}^{(1)}]^\dagger K_{xq}V_{\mathrm{rot}}^{(1)}.
    \label{eq:supp_recovery_scalar_residual}
\end{equation}
Its feature and derivative norms are
\begin{equation}
    \|\Delta_q\|_{\mathrm{op}}
    =\left|1-\frac{h_q}{a_q}\right||\sin\vartheta|,
    \qquad
    \|\partial_{\vartheta}\Delta_q\|_{\mathrm{op}}
    =\left|1-\frac{h_q}{a_q}\right||\cos\vartheta|.
    \label{eq:supp_recovery_scalar_norms}
\end{equation}
Thus, similar longitudinal and transverse responses support transfer of the scalar correction. At $\vartheta=0$, feature bias vanishes even when the derivative bias is nonzero. For finite regularization, the scalar score becomes $k_q=a_q/(a_q^2+\lambda_{\mathrm{reg}})$, and the corresponding residual is
\begin{equation}
    \Delta_{q,\lambda_{\mathrm{reg}}}
    =(k_qa_q-1)\cos\vartheta\,[V_{\mathrm{rot}}^{(1)}]^\dagger Z_qV_{\mathrm{rot}}^{(1)}
    +(1-k_qh_q)\sin\vartheta\,[V_{\mathrm{rot}}^{(1)}]^\dagger K_{xq}V_{\mathrm{rot}}^{(1)}.
    \label{eq:supp_recovery_regularized_residual}
\end{equation}
This example resolves the scalar attenuation picture into two response components and identifies the regularization contribution explicitly. Its closed-form expressions assume the stated two-layer CZ model, whereas the general residual identities above apply to other circuit architectures and depths. Resolving the transverse response through additional measurements of $K_{xq}$ and $K_{yq}=Y_q\prod_{q'\in\mathcal{N}(q)}Z_{q'}$ would define an augmented measurement protocol, beyond the computational-basis estimator used here.

\FloatBarrier

\section{Details of QNN training}\label{Method}

{This section specifies the numerical classifiers, noise parameters and calibration procedure used for Figs.~\ref{fig:loss}--\ref{fig:data2}; the separate protocol for Fig.~\ref{fig:data6}(c) is given in Sec.~\ref{app:training_comparison}. The details distinguish the illustrated CZ channel from the retained classification implementation.}

\subsection{Implementation overview}

{The software interfaces below implement the closed loop in Fig.~\ref{fig:fig1}(c), including the conversion from sampled bitstrings to fixed, calibrated features supplied to the classical model.}

The error-mitigated QNN training framework is implemented in Python. The classical components of the workflow, including data loading, batching, model construction, and optimization, are handled using \texttt{PyTorch} and \texttt{PyTorch-Lightning}~\cite{paszke2019pytorch}. Quantum circuits are constructed and simulated using \texttt{Qiskit}, while noisy circuit execution is performed with \texttt{Qiskit Aer} through the \texttt{AerSimulator} backend.

To integrate the quantum circuits into the hybrid learning pipeline, we use \texttt{Qiskit-Machine-Learning}. The quantum block is implemented using the \texttt{SamplerQNN} primitive and connected to the PyTorch model through the \texttt{TorchConnector}, allowing the quantum outputs to be incorporated into the end-to-end training loop. Classical data preprocessing, including feature scaling and train-validation-test splitting, is performed using \texttt{scikit-learn}~\cite{pedregosa2011scikit}. Reference fitting uses \texttt{SciPy}, and response relabelling and the linear solve are implemented in \texttt{NumPy}~\cite{virtanen2020scipy,harris2020numpy}.

\subsection{Data handling and preprocessing}\label{app:qnn_data}

{The splits and transformations here define the training and generalization tasks in Figs.~\ref{fig:loss}--\ref{fig:data2}; fitted preprocessing uses training data only.}

We evaluate the QNN classifiers using two benchmark datasets:
Breast Cancer and Digits. The Breast Cancer dataset contains
569 samples with 30 numerical features for binary classification,
while the Digits dataset comprises 1797 handwritten-digit images
of size $8\times8$, represented by 64 features for ten-class
classification. The Breast Cancer dataset is used in
Figs.~3(d--f) and 4, whereas Digits is used in
Figs.~3(a--c) and 5.

Each dataset is divided into training, validation, and test
subsets with proportions of 70\%, 10\%, and 20\%, respectively.
Stratified sampling preserves the class distributions across
all subsets. Input features are standardized using
\texttt{scikit-learn}'s \texttt{StandardScaler}, with the
mean and standard deviation determined exclusively from the
training set and subsequently applied to the validation and
test sets to prevent data leakage.

\subsection{Hybrid quantum-classical architecture}\label{app:qnn_architecture}

{The three stages below specify which quantities are optimized in the classifier and where the fixed estimator enters Fig.~\ref{fig:fig1}(c). They also determine the dimensions used in the main-text recovery equations.}

We employ a hybrid classifier that combines classical preprocessing and postprocessing layers with a parallel quantum circuit block, as shown in Fig.~\ref{fig:fig1}. This modular architecture allows the classical layers to adapt the input and output dimensions to the available quantum resources, while the quantum block supplies trainable measurement features for classification. The model consists of three stages.

\begin{enumerate}[leftmargin=*]
	\item \textbf{Classical preprocessing.}
	The standardized input feature vector $\mathbf{x}\in\mathbb{R}^{K}$ is first mapped to a latent representation by a trainable classical linear layer,
	\begin{equation}
		\mathbf{x}_{\mathrm{latent}}
		=
		W_{\mathrm{pre}}\mathbf{x}
		+
		\mathbf{b}_{\mathrm{pre}} .
		\label{eq:preprocessing}
	\end{equation}
	For a QNN with $P$ parallel quantum channels and $N$ qubits per channel, the latent vector has dimension $NP$ and is partitioned as
	\begin{equation}
\mathbf{x}_{\mathrm{latent}}
    ={\begin{bmatrix}\mathbf{x}^{(1)}\\ \vdots\\ \mathbf{x}^{(P)}\end{bmatrix}},
    \label{eq:latent_partition}
\end{equation}
	where $\mathbf{x}^{(\alpha)}\in\mathbb{R}^{N}$ denotes the feature subvector assigned to the $\alpha$-th quantum channel.
	
	\item \textbf{Quantum block.}
	The latent features are distributed across the $P$ quantum channels. Each channel $\alpha\in\{1,\ldots,P\}$ receives the subvector $\mathbf{x}^{(\alpha)}$ and processes it using an $N$-qubit parameterized circuit,
	\begin{equation}
		V\!\left(\mathbf{x}^{(\alpha)},\boldsymbol{\theta}^{(\alpha)}\right).
	\end{equation}
	The circuit output is measured in the computational basis to obtain raw bitstring probabilities. Under noisy execution, these probabilities are postprocessed using the structure-preserving mitigation procedure described in Sec.~\ref{sec:mitigation}, yielding mitigated local expectation values.
	
	\item \textbf{Classical postprocessing.}
	For each quantum channel, we use local Pauli-$Z$ expectation values as quantum features,
	\begin{equation}
\mathbf{s}^{(\alpha)}
    =\big[\langle Z^{(\alpha)}_1\rangle,\ldots,\langle Z^{(\alpha)}_N\rangle\big]^T.
    \label{eq:channel_features_methods}
\end{equation}
	The outputs from all channels are concatenated into a single feature vector,
	\begin{equation}
\mathbf{s}
    =\big[(\mathbf{s}^{(1)})^T,\ldots,(\mathbf{s}^{(P)})^T\big]^T.
    \label{eq:all_channel_features}
\end{equation}
	This vector is then passed through a final trainable classical linear layer to produce the classification logits,
	\begin{equation}
		\mathbf{l}
		=
		W_{\mathrm{post}}\mathbf{s}
		+
		\mathbf{b}_{\mathrm{post}} .
		\label{eq:logits}
	\end{equation}
	Here, $\mathbf{l}\in\mathbb{R}^{C}$, where $C$ is the number of output classes.
\end{enumerate}

\subsection{Quantum circuit design}\label{app:qnn_circuit}

{The gate choices and final rotation layers below specify the actual circuits behind the numerical classifiers, so they are needed to interpret circuit depth and calibration matching in Figs.~\ref{fig:loss} and~\ref{fig:data2}.}

Each quantum channel is implemented by a parameterized quantum circuit composed of a data-encoding block followed by a trainable variational ansatz,
\begin{equation}
V(\mathbf{x}^{(\alpha)},\boldsymbol{\theta}^{(\alpha)})
    =V_{\mathrm{train}}(\boldsymbol{\theta}^{(\alpha)})
    V_{\mathrm{enc}}(\mathbf{x}^{(\alpha)}).
    \label{eq:circuit_decomposition}
\end{equation}
Here, $\mathbf{x}^{(\alpha)}\in\mathbb{R}^{N}$ is the input assigned to channel $\alpha$, while the original sample input has dimension $K$. Channel indices are suppressed below when only one channel is considered.

\subsubsection{Data encoding}
For the $\alpha$-th quantum channel, the latent feature vector $\mathbf{x}^{(\alpha)}$ is encoded into an $N$-qubit state using single-qubit $R_y$ rotations. For a channel initialized in the reference state $\ket{0}^{\otimes N}$, the encoding unitary is
\begin{equation}
	V_{\mathrm{enc}}\!\left(\mathbf{x}^{(\alpha)}\right)
	=
	\bigotimes_{q=1}^{N}
	R_y\!\left(x_q^{(\alpha)}\right),
	\label{eq:data_encoding}
\end{equation}
where
\begin{equation}
	R_y(\phi)
	=
	\exp\!\left(-i\phi Y/2\right).
\end{equation}
Thus, each latent feature component $x_q^{(\alpha)}$ is encoded as a rotation angle on qubit $q$.

\subsubsection{Variational ansatz}

{The following ansatz identifies the entangling pattern and rotation-layer convention used by the retained classification runs; the CZ chain in Fig.~\ref{fig:fig1} remains an architectural illustration.}

The circuit implementations for Figs.~\ref{fig:loss} and \ref{fig:data2} use Qiskit \texttt{TwoLocal} with \texttt{rotation\_blocks=[ry,rz]}, \texttt{entanglement\_blocks=cx}, \texttt{entanglement=full}, and \texttt{reps=L}. The default final rotation layer is retained. Thus
\begin{equation}
V_{\mathrm{train}}(\boldsymbol{\theta})=
V_{\mathrm{rot}}^{(L+1)}
\prod_{\ell=1}^{L}\left[V_{\mathrm{ent,CX}}^{(\ell)}V_{\mathrm{rot}}^{(\ell)}\right],
\label{eq:trainable_ansatz}
\end{equation}
where $\ell=1$ acts first (rightmost), and the ordered CX pattern is the full-entanglement pattern generated by \texttt{TwoLocal}. The per-qubit rotations are
\begin{equation}
V_{\mathrm{rot}}^{(\ell)}=
\bigotimes_{q=1}^{N}R_z(\theta_{q,z}^{(\ell)})R_y(\theta_{q,y}^{(\ell)}).
\label{eq:rotation_layer}
\end{equation}
The number of trainable quantum angles per channel is $2N(L+1)$: 24 for $N=4,L=2$, and 40 for $N=4,L=4$. 

Figure~\ref{fig:fig1} and the closed-form example in Sec.~\ref{supp:observable_recovery} instead illustrate a CZ-chain circuit with $L$ rotation layers and no extra final rotation layer. Its entangler is
\begin{equation}
V_{\mathrm{ent,CZ}}=
\prod_{\substack{1\le q<N\\q\ \mathrm{odd}}}\mathrm{CZ}_{q,q+1}
\prod_{\substack{1\le q<N\\q\ \mathrm{even}}}\mathrm{CZ}_{q,q+1}.
\label{eq:entangling_layer}
\end{equation}

\subsection{Local noise model}\label{app:qnn_noise}

To evaluate QNN training under controlled noisy conditions, we use the Qiskit \texttt{AerSimulator} with a phenomenological noise model that captures the dominant error sources expected in superconducting-qubit processors~\cite{shen2026condensed}. The model includes single-qubit gate errors, two-qubit gate errors, readout errors, and finite-shot sampling noise.

\begin{enumerate}[leftmargin=*]
	\item \textbf{Single-qubit gate errors.}
	A single-qubit depolarizing channel with parameter $e_1$ can be applied after single-qubit gates.
	
	\item \textbf{Two-qubit gate errors.}
	After each entangling gate, we apply a two-qubit Pauli error channel. The simulation noise model attaches this channel to both CX and CZ instructions, according to the entangling gates used by each circuit. Specifically, the error is modeled as independent bit-flip errors on the two qubits participating in the gate. For a two-qubit state $\rho$, this channel takes the form
	\begin{equation}
		\mathcal{E}_{2}(\rho)
		=
		\sum_{a,b\in\{0,1\}}
		p_a p_b
		\left(X^a\otimes X^b\right)
		\rho
		\left(X^a\otimes X^b\right),
		\label{eq:two_qubit_noise}
	\end{equation}
	where $p_1=e_2$ and $p_0=1-e_2$. The parameter $e_2$ therefore controls the two-qubit gate noise strength. This parameter is the independent error probability on each endpoint, not the total probability that the two-qubit error is non-identity, which is $1-(1-e_2)^2$.
	
	\item \textbf{Readout errors.}
	Before measurement, we apply a classical readout error model. Each measured bit is flipped with probability $e_{\mathrm{ro}}$, corresponding to the assignment errors
	\begin{equation}
		{\rm Pr}(1|0) = {\rm Pr}(0|1) = e_{\mathrm{ro}} .
	\end{equation}
	Unless otherwise stated, we set
	\(
	e_{\mathrm{ro}} = 0.01 .
	\)
\end{enumerate}

\subsection{Details of the structure-preserving error mitigation}\label{sec:mitigation}

We employ a structure-preserving error-mitigation procedure designed to characterize and correct the effective noise associated with a fixed QNN circuit architecture. The key assumption is that, for a given device calibration window, the dominant noise affecting the measured output distribution is primarily determined by the circuit topology, including the gate set, connectivity, and layer structure, and depends only weakly on the specific trainable rotation angles $\boldsymbol{\theta}$. Under this assumption, a calibration performed once for a representative circuit with the same topology can be reused throughout QNN training.

\subsubsection{Identity fitting}

The first step is to construct a reference circuit $V_{\mathrm{mit}}(\boldsymbol{\theta})$ that has the same gate topology as the QNN circuit $V_{\mathrm{train}}(\boldsymbol{\theta})$.
To isolate the architecture-dependent noise, we optimize the parameters $\boldsymbol{\theta}$ such that the fitted circuit approximates the identity operation. Specifically, we determine
\begin{equation}
	\boldsymbol{\theta}_{\mathrm{id}}
	=
	\arg\min_{\boldsymbol{\theta}}
	\left\|
	V_{\mathrm{mit}}(\boldsymbol{\theta})
	-
	I
	\right\|_{F}^{2},
	\label{eq:identity_fitting}
\end{equation}
where $\|\cdot\|_F$ is the Frobenius norm. Reference fitting uses SciPy \texttt{L-BFGS-B} with all-zero initial angles, a maximum of 2000 iterations, and \texttt{ftol=$10^{-9}$}. The fitted unitary is characterized by its Frobenius-norm distance from the identity.

\subsubsection{Calibration matrix construction}

We next execute the checked reference to construct a raw response and relabel it as in Eq.~\eqref{eq:reference_relabel}. Let $\mathbf{p}^{\mathrm{ideal}}$ denote the ideal output probability distribution of an $N$-qubit circuit, written as a column vector in the computational basis. We model the noisy measured distribution as
\begin{equation}
	\mathbf{p}^{\mathrm{noisy}}
	\approx
	M_{\mathrm{mit}}
	\mathbf{p}^{\mathrm{ideal}},
	\label{eq:probability_noise_model}
\end{equation}
where $M_{\mathrm{mit}}$ is the effective response in ideal-output labels. This probability-level model is an approximation whose transfer residual is defined in Sec.~\ref{supp:observable_recovery}. Topology alone does not make it exact.

To estimate $M_{\mathrm{ref}}$, we prepare each state in the computational basis
$\{\ket{b}\}_{b=0}^{2^{N}-1}$ and execute the noisy implementation of
$V_{\mathrm{mit}}(\boldsymbol{\theta}_{\mathrm{id}})$ for $S_{\mathrm{calib}}$ shots. For each input bitstring $b$, we estimate the probability of measuring the output bitstring $b'$. The raw reference response is then
\begin{equation} {
		\big[M_{\mathrm{ref}}\big]_{b'b}
		=
		{\rm Pr}(b'|b)_{\mathrm{noisy}},}
	\label{eq:mmit_definition}
\end{equation}
where the column labels the prepared input and the row labels the measured output. The effective response is $M_{\mathrm{mit}}=M_{\mathrm{ref}}\Pi^T$, with $\Pi=I$ for a verified identity reference.

This construction does not require full quantum process tomography. It measures basis-input probabilities for the reference. Accurate correction of deployed circuits additionally requires a compatible effective response.

\subsubsection{Mitigated expectation-value estimation}

{The estimator is derived for the local features actually used by the classifier. A corrected quantum state or full output distribution is not needed for this step.} During QNN training, each quantum channel outputs local Pauli-$Z$ expectation values. For channel $\alpha$, the expectation value of $Z_q$ is
\begin{equation}
s_q^{(\alpha)}
    =\bra{\psi(\mathbf{x}^{(\alpha)},\boldsymbol{\theta}^{(\alpha)})}
      Z_q
      \ket{\psi(\mathbf{x}^{(\alpha)},\boldsymbol{\theta}^{(\alpha)})},
    \label{si}
\end{equation}
where
\begin{equation}
	\ket{\psi(\mathbf{x}^{(\alpha)},\boldsymbol{\theta}^{(\alpha)})}
	=
	V(\mathbf{x}^{(\alpha)},\boldsymbol{\theta}^{(\alpha)})
	\ket{\psi_0},
	\qquad
	\ket{\psi_0}=\ket{0}^{\otimes N}.
\end{equation}

These expectation values are linear functions of the output probability distribution. To express this relation, we define the measurement functional matrix
$F\in\{-1,+1\}^{2^N\times N}$, whose element $F_{bq}$ is the eigenvalue of $Z_q$ on the computational-basis state $\ket{b}$:
\begin{equation}
	F_{bq}
	=
	\bra{b}
	Z_q
	\ket{b}.
	\label{eq:F_definition}
\end{equation}
For an ideal probability distribution $\mathbf{p}^{\mathrm{ideal}}$, the corresponding vector of ideal local magnetizations is
\begin{equation}
	\mathbf{s}^{\mathrm{ideal}}
	=
	F^{T}
	\mathbf{p}^{\mathrm{ideal}}.
	\label{eq:ideal_expectation}
\end{equation}

A direct mitigation approach would estimate
$\mathbf{p}^{\mathrm{ideal}}\approx M_{\mathrm{mit}}^{-1}\mathbf{p}^{\mathrm{noisy}}$
and then compute
$\mathbf{s}^{\mathrm{mit}}=F^T M_{\mathrm{mit}}^{-1}\mathbf{p}^{\mathrm{noisy}}$.
However, direct inversion of $M_{\mathrm{mit}}$ can be numerically unstable, especially when the matrix is ill-conditioned or affected by finite-shot calibration noise. Therefore, we instead construct a regularized linear estimator $\eta$ that maps noisy probabilities directly to mitigated expectation values:
\begin{equation}
	\mathbf{s}^{\mathrm{mit}}
	=
	\eta^{T}
	\mathbf{p}^{\mathrm{noisy}}.
	\label{eq:eta_estimator}
\end{equation}

The estimator $\eta$ is chosen so that Eq.~\eqref{eq:eta_estimator} reproduces the ideal expectation values whenever the noisy distribution satisfies Eq.~\eqref{eq:probability_noise_model}. Substituting
$\mathbf{p}^{\mathrm{noisy}}\approx M_{\mathrm{mit}}\mathbf{p}^{\mathrm{ideal}}$, we require $\mathbf{s}^{\mathrm{mit}} \approx \mathbf{s}^{\mathrm{ideal}}$:
\begin{equation}
	\eta^{T}
	M_{\mathrm{mit}}
	\mathbf{p}^{\mathrm{ideal}}
	\approx
	F^{T}
	\mathbf{p}^{\mathrm{ideal}},
\end{equation}
for arbitrary $\mathbf{p}^{\mathrm{ideal}}$, which gives the linear condition
\begin{equation}
	M_{\mathrm{mit}}^{T}\eta
	\approx
	F.
\end{equation}
We solve this system with Tikhonov regularization by minimizing
\begin{equation}
	\min_{\eta}
	\left\|
	M_{\mathrm{mit}}^{T}\eta
	-
	F
	\right\|_{F}^{2}
	+
	\lambda_{\mathrm{reg}}
	\left\|
	\eta
	\right\|_{F}^{2},
	\label{eq:tikhonov_objective}
\end{equation}
where $\lambda_{\mathrm{reg}}$ is a small regularization parameter. In our simulations, we use $\lambda_{\mathrm{reg}}=10^{-6}$. The closed-form solution is
\begin{equation}{
		\eta
		=
		\left(
		M_{\mathrm{mit}}M_{\mathrm{mit}}^{T}
		+
		\lambda_{\mathrm{reg}} I
		\right)^{-1}
		M_{\mathrm{mit}}F.}
	\label{eq:eta_solution}
\end{equation}

During the forward pass of the QNN, the same precomputed estimator $\eta$ is applied independently to each parallel quantum channel. For channel $\alpha$, the noisy measured probability vector $[\mathbf{p}^{(\alpha)}]^{\mathrm{noisy}}$ is mapped to mitigated expectation values as
\begin{equation}
	[\mathbf{s}^{(\alpha)}]^{\mathrm{mit}}
	=
	\eta^{T}
	[\mathbf{p}^{(\alpha)}]^{\mathrm{noisy}}.
	\label{mits}
\end{equation}
The mitigated vectors from all channels are then concatenated and passed to the classical postprocessing layer. In this way, the QNN is trained using corrected quantum features while requiring only a single reusable calibration for the fixed circuit architecture.

\subsection{Training procedure}\label{app:qnn_training}

The hybrid QNN is trained end to end using the \texttt{PyTorch-Lightning} framework. The trainable parameters include the classical preprocessing parameters $W_{\mathrm{pre}}$ and $\mathbf{b}_{\mathrm{pre}}$, the classical postprocessing parameters $W_{\mathrm{post}}$ and $\mathbf{b}_{\mathrm{post}}$, and the quantum variational parameters $\boldsymbol{\theta}$. We collect all trainable parameters as
\begin{equation}
	\boldsymbol{v}
	=
	\{
	W_{\mathrm{pre}},
	\mathbf{b}_{\mathrm{pre}},
	W_{\mathrm{post}},
	\mathbf{b}_{\mathrm{post}},
	\boldsymbol{\theta}
	\}.
	\label{eq:trainable_parameters}
\end{equation}

The predicted class probabilities are obtained from the logits $\mathbf{l}$ in Eq.~\eqref{eq:logits} using the softmax function,
\begin{equation}
	f_k
	=
	\frac{\exp(l_k)}
	{\sum_{k'=1}^{C}\exp(l_{k'})},
	\qquad
	k=1,\ldots,C .
	\label{eq:softmax}
\end{equation}
Here, $C$ is the number of classes and $f_k$ denotes the predicted probability for class $k$.

For a dataset containing $Q$ samples, the $i$-th sample is represented by an input feature vector $\mathbf{x}_i\in\mathbb{R}^{K}$ and a one-hot label vector $\mathbf{y}_i\in\{0,1\}^{C}$, with components $y_{i,k}$. The corresponding predicted probability vector is $\mathbf{f}_i=[f_{i,1},\ldots,f_{i,C}]^T$. We train the model by minimizing the empirical multiclass cross-entropy loss
\begin{equation}
	\mathcal{L}(\boldsymbol{v})
	=
	-
	\frac{1}{Q}
	\sum_{i=1}^{Q}
	\sum_{k=1}^{C}
	y_{i,k}
	\ln
	f_{i,k}(\mathbf{x}_i;\boldsymbol{v}) .
	\label{eq:loss_dataset}
\end{equation}
Here, $f_{i,k}(\mathbf{x}_i;\boldsymbol{v})$ is the predicted probability of class $k$ for input $\mathbf{x}_i$, evaluated using the full set of trainable parameters $\boldsymbol{v}$.

The forward pass of the error-mitigated QNN proceeds as follows. First, the standardized input data are transformed by the classical preprocessing layer and distributed across the parallel quantum channels. Second, each quantum channel is executed and its noisy output probabilities are converted into mitigated expectation values using the precomputed estimator in Eq.~\eqref{mits}. Unmitigated runs use the raw expectation values instead. Third, the quantum features are concatenated and passed through the classical postprocessing layer to obtain the logits and predicted class probabilities. Finally, the loss in Eq.~\eqref{eq:loss_dataset} is evaluated and used to update both the quantum and classical parameters.

Optimization is performed using the AdamW optimizer~\cite{loshchilov2019decoupled}. We use AdamW because its decoupled weight-decay update provides a stable optimization procedure for hybrid models containing both classical neural-network weights and quantum variational parameters.

\subsection{Variational-training comparison}\label{app:training_comparison}

The new local benchmark in Fig.~\ref{fig:data6}(c) minimizes the normalized mean magnetization in Eq.~\eqref{z}. It uses five qubits initialized in $\ket{0}^{\otimes5}$ and four layers, each consisting of per-qubit $R_y$ rotations followed by $R_z$ rotations and then CZ gates on the even and odd nearest-neighbor bonds. There are 40 trainable angles and 16 CZ gates, with no additional final rotation layer. After every CZ gate, each endpoint independently undergoes a Pauli-$X$ error with probability 0.006; each measured bit is independently flipped with probability 0.01. Single-qubit rotation gates are otherwise noiseless. The noiseless control omits both error channels but retains the same finite-shot sampling. We perform five independent finite-shot repeats using all-zero initial angles, 1,000 shots per executed circuit, and COBYLA~\cite{powell1994direct} with an initial trust-region radius of 1, a tolerance of $10^{-8}$, and a maximum of 500 objective evaluations. Thus, the horizontal axis counts objective evaluations rather than gradient updates or auxiliary circuits.

Readout mitigation uses a measured $32\times32$ basis-assignment matrix. Our method instead measures the same basis inputs through the noisy four-layer reference with all rotation angles zero. Its ideal action is verified to be the identity. Both calibrations use 4,000 shots per basis input and the regularized feature estimator in Eq.~\eqref{eq:eta_solution}, with $\lambda_{\mathrm{reg}}=10^{-6}$. The resulting estimator is fixed during a training run. ZNE uses local CZ folding with factors 1, 3, and 5, leaving ideal circuit action unchanged, and quadratic Richardson weights $15/8$, $-5/4$, and $3/8$ \cite{temme2017error,li2017efficient}. Readout noise is not amplified or separately corrected in this ZNE condition.

At every target evaluation, CDR fits a scalar affine map from 32 noisy and noiseless near-Clifford auxiliary magnetizations \cite{czarnik2021error}. Each auxiliary circuit retains eight randomly chosen target rotation angles; the remaining rotations are stochastically replaced by multiples of $\pi/2$, favoring nearby gates with weights $\exp(-d^2/0.7^2)$, where $d$ is their phase-insensitive Frobenius distance. The slope is fitted using centered ridge regression with a penalty of $10^{-6}$ and an unpenalized intercept~\cite{hoerl1970ridge}. No exact target-circuit objective is supplied to CDR or the optimizer. Neither CDR nor ZNE is stacked with separate readout mitigation in this six-condition comparison. Corrected objectives are not clipped to the physical magnetization range.

At 500 objective evaluations, the simulated measurement budgets per repeat are 500,000 shots for noiseless and raw noisy training, 628,000 shots for both readout mitigation and our method (including calibration), 1,500,000 shots for ZNE, and 16,500,000 shots for this local CDR implementation. Equal evaluation counts therefore do not imply equal shot budgets. Exact density-matrix evolution supplies circuit probabilities, from which multinomial bitstring counts are sampled. The simulator is independently checked against Qiskit Aer density-matrix execution with explicit Pauli channels. Ideal evaluations of the trained states are retained only as diagnostics and are not used for parameter updates or selection. The optimizer selects parameters using its own training estimator. All new results in panel (c) are local simulations; panels (a,b) retain their existing source data.

\section{IBM Quantum device}

{The device information below locates the hardware demonstration in Fig.~\ref{fig:data6}(b). Note that this is distinct from the explicit noise channels used in the controlled simulator comparisons.}

\label{app:ibm}
\begin{figure}[ht]
    \centering
    \includegraphics[width=0.8\linewidth]{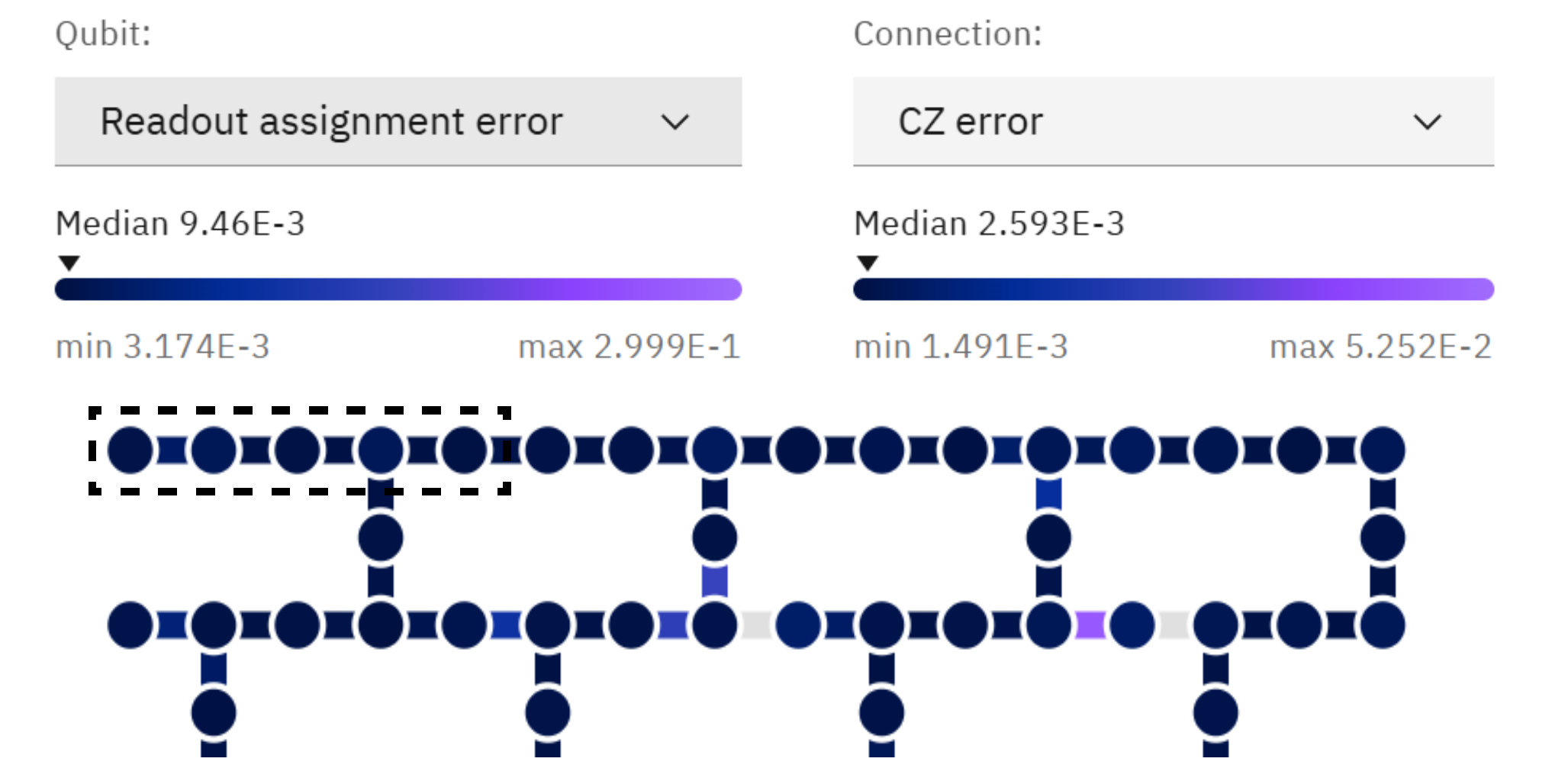}
    \caption{Noise data for the \texttt{ibm\_fez} device. The selected qubit chain is marked by the dashed box. This device information is separate from the phenomenological CX/CZ noise model used for the controlled QNN training benchmarks.}
    \label{fig:ibm}
\end{figure}

Figure~\ref{fig:ibm} shows the hardware noise information for the \texttt{ibm\_fez} device, with the selected qubit chain marked by the dashed box. The controlled QNN simulations in Figs.~\ref{fig:loss}--\ref{fig:data2} instead use the explicit phenomenological model and settings given in Sec.~\ref{Method}.

\section{Noise effects in circuit optimization}
\label{app:noise_effects_optimization}

{This appendix gives a contraction picture for interpreting the optimization results in Fig.~\ref{fig:data6} and the schematic feature contraction and correction in Fig.~\ref{fig:fig1}(d).}  We consider an $N$-qubit parameterized circuit $V_{\rm train}(\boldsymbol{\theta})$, where $\boldsymbol{\theta}$ denotes the trainable circuit parameters. The circuit acts on the reference state
\begin{equation}
    \rho_0
    =
    \bigl(\ket{0}\bra{0}\bigr)^{\otimes N}.
\end{equation}
The ideal output state is
\begin{equation}
    \rho^{\rm ideal}(\boldsymbol{\theta})
    =
    V_{\rm train}(\boldsymbol{\theta})
    \rho_0
    V_{\rm train}^{\dagger}(\boldsymbol{\theta}).
\end{equation}

We first consider the variational benchmark used in the main text, where the objective is the normalized average Pauli-$Z$ magnetization,
\begin{equation}
    \mathcal{C}_{Z}^{\rm ideal}(\boldsymbol{\theta})
    =
    \frac{1}{N}
    \sum_{q=1}^{N}
    {\rm Tr}
    \left[
    Z_q
    \rho^{\rm ideal}(\boldsymbol{\theta})
    \right],
    \label{eq:app_cz_ideal}
\end{equation}
where $Z_q$ is the Pauli-$Z$ operator acting on qubit $q$.  In a noisy implementation, the ideal state is transformed by an effective noise channel $\mathcal{N}$ associated with the chosen circuit architecture,
\begin{equation}
    \rho^{\rm noisy}(\boldsymbol{\theta})
    =
    \mathcal{N}
    \left[
    \rho^{\rm ideal}(\boldsymbol{\theta})
    \right].
\end{equation}
The corresponding noisy objective is
\begin{equation}
    \mathcal{C}_{Z}^{\rm noisy}(\boldsymbol{\theta})
    =
    \frac{1}{N}
    \sum_{q=1}^{N}
    {\rm Tr}
    \left[
    Z_q
    \rho^{\rm noisy}(\boldsymbol{\theta})
    \right].
    \label{eq:app_cz_noisy_def}
\end{equation}

Because quantum channels are linear maps on density matrices and expectation values are linear functions of the state, the noisy objective is a linear response of the ideal output information.  When the dominant effect of noise in the measured $Z$ sector is signal attenuation, this response may be approximated by a scalar contraction,
\begin{equation}
    \mathcal{C}_{Z}^{\rm noisy}(\boldsymbol{\theta})
    \simeq
    \lambda_Z
    \mathcal{C}_{Z}^{\rm ideal}(\boldsymbol{\theta})
    +
    \xi_Z,
    \qquad
    0<\lambda_Z<1.
    \label{eq:app_cz_noise_contraction}
\end{equation}
Here, $\lambda_Z$ denotes the effective noise-induced contraction of the measured $Z$ signal, while $\xi_Z$ represents finite-shot fluctuations and residual hardware fluctuations.  For two parameter points $\boldsymbol{\theta}$ and $\boldsymbol{\theta}'$, define
\begin{equation}
    \Delta \mathcal{C}_{Z}^{\rm ideal}
    =
    \mathcal{C}_{Z}^{\rm ideal}(\boldsymbol{\theta})
    -
    \mathcal{C}_{Z}^{\rm ideal}(\boldsymbol{\theta}'),
    \qquad
    \Delta \mathcal{C}_{Z}^{\rm noisy}
    =
    \mathcal{C}_{Z}^{\rm noisy}(\boldsymbol{\theta})
    -
    \mathcal{C}_{Z}^{\rm noisy}(\boldsymbol{\theta}').
\end{equation}
If the fluctuation term is averaged over or is small compared with the systematic contraction, Eq.~\eqref{eq:app_cz_noise_contraction} gives
\begin{equation}
    \Delta \mathcal{C}_{Z}^{\rm noisy}
    \simeq
    \lambda_Z
    \Delta \mathcal{C}_{Z}^{\rm ideal}.
    \label{eq:app_cz_cost_difference}
\end{equation}
Similarly, the gradient is suppressed as
\begin{equation}
    \partial_{\boldsymbol{\theta}}
    \mathcal{C}_{Z}^{\rm noisy}
    \simeq
    \lambda_Z
    \partial_{\boldsymbol{\theta}}
    \mathcal{C}_{Z}^{\rm ideal}.
    \label{eq:app_cz_gradient_contraction}
\end{equation}
{Within this parameter-independent contraction approximation, noise reduces the mean cost difference and its parameter derivative. This explains the flattening mechanism without asserting it for every circuit or noise channel.}

The same reasoning applies to the QNN feature vector of one channel (the channel index is suppressed). Let
\begin{equation}
    \mathbf{s}^{\rm ideal}(\boldsymbol{\theta})
    =
    \left(
    s_1^{\rm ideal},
    s_2^{\rm ideal},
    \ldots,
    s_N^{\rm ideal}
    \right)^T
\end{equation}
be the vector of ideal local Pauli-$Z$ features, with
\begin{equation}
    s_q^{\rm ideal}(\boldsymbol{\theta})
    =
    {\rm Tr}
    \left[
    Z_q
    \rho^{\rm ideal}(\boldsymbol{\theta})
    \right].
\end{equation}
Under noisy execution,
\begin{equation}
    s_q^{\rm noisy}(\boldsymbol{\theta})
    =
    {\rm Tr}
    \left[
    Z_q
    \mathcal{N}
    \left[
    \rho^{\rm ideal}(\boldsymbol{\theta})
    \right]
    \right].
    \label{eq:app_noisy_feature_trace}
\end{equation}
{When the noisy adjoint response approximately closes on the measured local Pauli-$Z$ observables, its action can be represented within that subspace by an effective linear map,}
\begin{equation}
    \mathbf{s}^{\rm noisy}
    \simeq
    A_{\rm noise}
    \mathbf{s}^{\rm ideal}
    +
    \boldsymbol{\xi},
    \label{eq:app_feature_linear_map}
\end{equation}
where $A_{\rm noise}$ is the effective feature-space noise response and $\boldsymbol{\xi}$ denotes finite-shot and hardware fluctuations. {Outside this approximate closure, the probability/observable residuals of Appendix~\ref{supp:observable_recovery} are needed. The additional assumption of nearly equal local contractions gives $A_{\rm noise}\simeq\lambda I$ and hence}
\begin{equation}
    \mathbf{s}^{\rm noisy}
    \simeq
    \lambda
    \mathbf{s}^{\rm ideal}
    +
    \boldsymbol{\xi},
    \qquad
    0<\lambda<1.
    \label{eq:app_noisy_feature_simple}
\end{equation}
If $\lambda$ is mainly determined by the circuit architecture and device noise and depends only weakly on the variational parameters, then
\begin{equation}
    \partial_{\boldsymbol{\theta}}
    \mathbf{s}^{\rm noisy}
    \simeq
    \lambda
    \partial_{\boldsymbol{\theta}}
    \mathbf{s}^{\rm ideal}.
    \label{eq:app_noisy_feature_gradient_simple}
\end{equation}{Equation~\eqref{eq:app_noisy_feature_gradient_simple} describes suppression of the feature Jacobian in this approximation. Its effect on the loss also depends on the classical readout, as discussed for Fig.~\ref{fig:fig1}(d). A flatter landscape and slower convergence are possible consequences, rather than conclusions for an arbitrary QNN or noise channel.}

We now describe the mitigation effect in the same language.  The structure-preserving mitigation protocol learns an architecture-specific correction from a reference circuit with the same gate topology as the QNN block.  {The implemented correction solves constraints for the measured features rather than inverting a full quantum noise process. If its action is approximately a scalar rescaling of the measured magnetization, it can be written phenomenologically as}
\begin{equation}
    \mathcal{C}_{Z}^{\rm mit}(\boldsymbol{\theta})
    \simeq
    \lambda_{\rm mit}^{-1}
    \mathcal{C}_{Z}^{\rm noisy}(\boldsymbol{\theta}),
    \label{eq:app_cz_mitigation_rescale}
\end{equation}
where $\lambda_{\rm mit}^{-1}$ is the effective inverse signal loss learned from calibration.  Consequently,
\begin{equation}
    \Delta \mathcal{C}_{Z}^{\rm mit}
    \simeq
    \lambda_{\rm mit}^{-1}
    \Delta \mathcal{C}_{Z}^{\rm noisy},
    \qquad
    \partial_{\boldsymbol{\theta}}
    \mathcal{C}_{Z}^{\rm mit}
    \simeq
    \lambda_{\rm mit}^{-1}
    \partial_{\boldsymbol{\theta}}
    \mathcal{C}_{Z}^{\rm noisy}.
    \label{eq:app_cz_mitigation_gradient}
\end{equation}
{When the calibrated scalar response matches the deployed one, these relations recover the corresponding mean signal differences and derivatives. A mismatch leaves the transfer bias quantified in Appendix~\ref{supp:observable_recovery}.}

For the same single-channel feature vector, the analogous expression is
\begin{equation}
    \mathbf{s}^{\rm mit}
    \simeq
    A_{\rm mit}^{-1}
    \mathbf{s}^{\rm noisy},
    \label{eq:app_feature_mitigation_general}
\end{equation}
where $A_{\rm mit}^{-1}$ denotes the effective inverse map obtained from the structure-preserving calibration.  In the dominant-contraction approximation,
\begin{equation}
    \mathbf{s}^{\rm mit}
    \simeq
    \lambda_{\rm mit}^{-1}
    \mathbf{s}^{\rm noisy}.
    \label{eq:app_feature_mitigation_simple}
\end{equation}
In practice, this inverse is not implemented as an unstable direct inversion of the calibration matrix. Instead, the protocol uses the regularized feature estimator of Eq.~\eqref{eq:eta_solution}, which corrects the measured Pauli-$Z$ features while suppressing noise-sensitive directions.

% \section{Perturbative effect}
% For a variational circuit V, given the ideal outcome $\ket{\psi}=V\ket{\psi_{0}}$, the perturbed state is thus defined as
% \begin{equation}
%         \ket{\psi_{p}}=V_{p}V\ket{\psi_{0}}
% \end{equation}
% The perturbative block can be expressed as a product of two-qubit entangling operations with slightly shifted parameters:
% \begin{equation}
% V=\prod_{i\in even} R^{zz}_{i}  \prod_{i\in odd} R^{zz}_{i},
% \end{equation}
% where $R^{zz}=\exp(-i\delta \theta ZZ)$, and $\delta \theta$ is for a small random number.

% In the QNN implementation, we first obtain the ideal output state $\ket{\psi}$ from noiseless circuit simulations. To emulate perturbative or noisy effects, we then numerically apply a perturbation operator over this state, leading to  $V_{p}\ket{\psi}$.

\clearpage
\section{Comparison with other mitigation methods in QNN training}

 {The comparison below separates reference-calibration cost from the circuits required at each new parameter point. It explains the operational role of reuse in Figs.~\ref{fig:data6}(c) and~\ref{fig:qnncompare} without claiming that reuse is unique to this protocol.}

\label{supp:mitigation_comparison}

The practical advantage of the present method is that calibration is performed for an elementary circuit block and its feature estimator is then used inside the training loop. Setup comprises classical identity fitting, $2^N$ basis-input reference experiments, and a regularized linear solve [Sec.~\ref{sec:mitigation}]. The four-qubit channels require a $16\times16$ response and a $16\times4$ estimator. The shot allocations in Figs.~\ref{fig:noisyqnn} and \ref{fig:data2} correspond to 64,000 and 160,000 shots per reference calibration, respectively. When the effective responses are compatible, this setup is shared across the $P$ channels and subsequent parameter updates; increasing network width does not enlarge the shared calibration matrix. During training, the quantum and classical model parameters are updated while $\eta$ remains fixed. For channel $\alpha$, the corrected features are 
\begin{equation}
	[\mathbf{s}^{(\alpha)}]^{\mathrm{mit}}
	=\eta^T[\mathbf{p}^{(\alpha)}]^{\mathrm{noisy}}.
	\label{eq:supp_training_correction}
\end{equation}
The same estimator can act on each channel.

Readout mitigation can likewise retain an assignment calibration while circuit parameters change. Its correction targets measurement errors and leaves gate errors in the underlying distribution \cite{nation2021scalable,vandenberg2022model}. Our reference instead executes the channel's gate pattern, so the calibrated response includes accumulated gate and readout effects. This extends calibration to circuit-induced feature distortions, subject to the transfer conditions above, rather than relying on detector characterization alone.

Conventional pointwise zero-noise extrapolation (ZNE) obtains each corrected estimate from measurements at several noise strengths \cite{temme2017error,li2017efficient,giurgicatiron2020digital,cai2021multiexponential}. During training, updated feature or derivative circuits therefore require corresponding noise-scaled data; retaining extrapolation coefficients does not supply these measurements. Sampled probabilistic error cancellation (PEC) instead estimates observables using weighted outcomes from corrective circuit instances \cite{endo2018practical}. Its noise models may be reused, but characterization can recur: the sparse-PEC experiment of Ref.~\cite{vandenberg2023probabilistic} periodically relearned its models to track temporal fluctuations. Such recalibration is distinct from measuring the current target circuits. Our implementation applies a fixed classical feature map to ordinary channel outputs, avoiding noise-scaled and corrective circuit families during compatible training updates.

Clifford data regression (CDR) constructs noisy/classically computed training pairs and fits a correction near a target circuit \cite{czarnik2021error}. Transferable corrections, including learned mitigation schemes, can remain useful across parameter changes \cite{czarnik2021error,strikis2021learning}. If a target-specific fit becomes invalid during optimization, updating it can require additional auxiliary measurements and classical target calculations. Here, the reference labels are known Pauli eigenvalues, and $\eta$ follows from the regularized response equation; neither near-Clifford training-set selection nor an auxiliary neural model is required~\cite{liao2025noise}. This provides a direct calibration procedure for the features used by the classifier.  Learned mitigation can also be trained across variational parameters \cite{korolev2026machine}, and real-time mitigation can retain a correction until a monitored response changes \cite{robbiati2026realtime}. The present choice differs in fitting a fixed linear estimator directly to basis-input measurement constraints for a repeated QNN block.

Standard virtual distillation uses multiple copies of each evaluated state \cite{huggins2021virtual,koczor2021exponential}. Symmetry verification requires a known symmetry sector and obtains the relevant checks from compatible outcomes or additional measurements; postselection discards outcomes outside that sector \cite{bonetmonroig2018lowcost,mcardle2019stabilizer,sagastizabal2019experimental}. The present feature estimator requires neither multiple copies nor a prescribed symmetry sector.

The demonstrated benefit is improved learning with this reusable block calibration. Figures~\ref{fig:noisyqnn} and \ref{fig:data2} show faster loss reduction and lower training losses than unmitigated training. In the 25-channel, 100-qubit example, mean test accuracy increases from 82.0\% to 83.6\% at $e_2=0.02$ and from 81.6\% to 83.6\% at $e_2=0.10$ [Fig.~\ref{fig:qnncompare}]. These results support the usefulness of direct feature correction throughout modular QNN training without requiring calibration of the full network.

\flushbottom
\end{document}